\documentclass[aip,groupedaddress,superscriptaddress,reprint]{revtex4-1}
\usepackage[bookmarks=true,colorlinks,linkcolor=OrangeRed,urlcolor=NavyBlue,citecolor=RoyalBlue]{hyperref}
\usepackage{amsmath,amssymb,color,comment,physics}
\usepackage[makeroom]{cancel}
\usepackage[caption=false]{subfig}
\usepackage{mathrsfs}
\usepackage{graphicx}
\usepackage{subfig}
\usepackage[countmax]{subfloat}
\usepackage[english]{babel}
\usepackage[dvipsnames]{xcolor}
\usepackage{mathtools}
\usepackage{MnSymbol}
\DeclareUnicodeCharacter{2212}{\textendash}

\usepackage{braket}

\definecolor{mypurple}{rgb}{0.49,0.18,0.56}
\definecolor{mygold}{rgb}{0.93,0.49,0.13}
\definecolor{mygreen}{rgb}{0,0.5,0}
\definecolor{myblue}{rgb}{0,0,0.75}
\definecolor{mymagenta}{cmyk}{0,1,0,0.12}
\definecolor{mygray}{rgb}{0.5,0.5,0.5}

\usepackage{umoline}

\definecolor{mypink1}{rgb}{0.858, 0.188, 0.478}

\begin{document}

\title{Short-time Accuracy and Intra-electron Correlation for Nonadiabatic Quantum-Classical Mapping Approaches}

\author{Haifeng Lang}
\thanks{Current address: Department of Nuclear Engineering and Management, Graduate School of Engineering, The University of Tokyo, 7-3-1 Hongo, Bunkyo-ku, Tokyo 113-8656, Japan}
\email{hflang@g.ecc.u-tokyo.ac.jp}
\affiliation{Pitaevskii BEC Center, CNR-INO and Dipartimento di Fisica, Universit\`a di Trento, Via Sommarive 14, Trento, I-38123, Italy}
\affiliation{Theoretical Chemistry, Institute of Physical Chemistry, Heidelberg University, Im Neuenheimer Feld 229, 69120 Heidelberg, Germany }

\author{Philipp Hauke}
\email{philipp.hauke@unitn.it }
\affiliation{Pitaevskii BEC Center, CNR-INO and Dipartimento di Fisica, Universit\`a di Trento, Via Sommarive 14, Trento, I-38123, Italy}
\affiliation{INFN-TIFPA, Trento Institute for Fundamental Physics and Applications, Trento, Italy}





\begin{abstract}

Nonadiabatic quantum-classical mapping approaches have significantly gained in popularity in the past several decades because they have acceptable accuracy while remaining numerically tractable even for large system sizes. In the recent few years, several novel mapping approaches have been developed that display higher accuracy than the traditional Ehrenfest method, linearized semiclassical initial value representation (LSC-IVR), and Poisson bracket mapping equation (PBME) approaches. While various benchmarks have already demonstrated the advantages and limitations of those methods, rigorous theoretical justifications of their short-time accuracy are still demanded. In this article, we systematically examine the intra-electron correlation, as a statistical measure of electronic phase space, which has been first formally proposed for mapping approaches in the context of the generalized discrete truncated Wigner approximation (GDTWA) and which is a key ingredient for the improvement of short-time accuracy of such mapping approaches. We rigorously establish the connection between short-time accuracy and intra-electron correlation for various widely used models. We find that LSC-IVR, PBME, and Ehrenfest methods fail to correctly reproduce the intra-electron correlation. While some of the traceless Meyer--Miller--Stock--Thoss (MMST) approaches, partially linearized density matrix (PLDM) approach, and spin partially linearized density matrix (Spin-PLDM) approach are able to sample the intra-electron correlation correctly, the spin linearized semiclassical (Spin-LSC) approach and the other traceless MMST approaches sample the intra-correlation faithfully only for two-level systems. Our theoretical analysis provides insights into the short-time accuracy of semiclassical methods and presents mathematical justifications for previous numerical benchmarks.

\end{abstract}

\pacs{Valid PACS appear here}
\maketitle
\section{Introduction}

Mapping approaches (linearized phase space methods) are powerful tools for the simulation of nonadiabatic processes in large scale condensed phase systems because of their balance between manageable resource requirements and reliable accuracy\cite{hillery1984distribution,steel1998dynamical,blakie2008dynamics,polkovnikov2010phase,schachenmayer2015many,zhu2019generalized,davidson2015s,wurtz2018cluster,polkovnikov2003quantum,orioli2017nonequilibrium,pucci2016simulation,meyera1979classical,cotton2013symmetrical,stock1997semiclassical,cotton2013symmetrical2,liu2017isomorphism,he2019new,liu2016unified,miller2017classical,saller2019identity,saller2019improved,sun1998semiclassical,kim2008quantum,kelly2012mapping,huo2011communication,huo2013communication,huo2012consistent,hsieh2012nonadiabatic,hsieh2013analysis,kapral1999mixed,stock1999flow,muller1999flow,cotton2015symmetrical,meyer1979classical,mannouch2020partially,mannouch2020partially2,runeson2019spin,runeson2020generalized,he2021commutator,he2022new,bossion2022non}. In such approaches, the quantum dynamics is approximated as classical dynamics of the phase space. The phase space for the nuclear degrees of freedom are naturally chosen as ordinary Wigner distributions, while different mapping approaches have different electronic phase space choices, such as the phase space of Meyer--Miller--Stock--Thoss (MMST) mapping harmonic oscillators\cite{meyera1979classical,cotton2013symmetrical,stock1997semiclassical,cotton2013symmetrical2,miller2017classical,saller2019identity,saller2019improved,sun1998semiclassical,kim2008quantum,kelly2012mapping,huo2011communication,huo2013communication,hsieh2012nonadiabatic,hsieh2013analysis,kapral1999mixed,stock1999flow,muller1999flow,huo2012consistent}, continuous\cite{cotton2015symmetrical,meyer1979classical,mannouch2020partially,mannouch2020partially2,runeson2019spin,runeson2020generalized} or discrete\cite{schachenmayer2015many,zhu2019generalized,lang2021generalized} spin phase space, phase space of angle-action variables\cite{cotton2013symmetrical,cotton2013symmetrical2,cotton2015symmetrical}, constrained phase space\cite{liu2016unified,liu2017isomorphism,he2019new,he2021commutator,he2022new,liu2021unified}, etc. The classical dynamics makes the necessary computing resources scale only linear with the system size. As a comparison, the required resources of the numerical converged methods increase exponentially\cite{manthe1992wave,meyer1990multi,beck2000multiconfiguration}. One major drawback of the mapping approaches, however, is that they cannot capture the nuclear quantum coherence correctly. Nevertheless, this limitation usually does not harm the achieved accuracy too much for the condensed phase systems\cite{gao2020benchmarking}. 
In the recent few years, various advanced mapping approaches\cite{cotton2013symmetrical,cotton2013symmetrical2,lang2021generalized,mannouch2020partially,mannouch2020partially2,runeson2019spin,runeson2020generalized,he2019new,huo2012consistent,runeson2020generalized,runeson2019spin,saller2019identity,saller2019improved,cheng2024novel,he2024nonadiabatic,wu2024nonadiabatic} have been proposed. Compared to the traditional Ehrenfest\cite{domcke2004conical}, Poisson bracket mapping equation\cite{kim2008quantum,kelly2012mapping,nassimi2010analysis} (PBME), and linearized semiclassical initial value representation\cite{sun1998semiclassical} (LSC-IVR), these methods provide a significant improvement in accuracy but without  seriously increasing the additional computation efforts. The reason is that these methods choose more suitable electronic phase spaces as compared to the traditional methods. Despite these advances, a rigorous connection between electronic phase space property and short-time accuracy of mapping approaches is still missing. 

In this work, such connection is established with the help of intra-electron correlation\cite{golosov2001classical,zhu2019generalized,lang2021generalized}. Intra-electron correlation is a measure of the statistical feature of the electronic phase space that has first been formally proposed in the context of the generalized discrete truncated Wigner approximation (GDTWA). In the original GDTWA paper\cite{zhu2019generalized}, the correct intra-electron correlation sampling states that the statistical average of the quadratic electronic (spin) phase space variables should be identical to the quantum expectation values of the symmetrized product of the corresponding electronic operators. A similar idea\cite{golosov2001classical} was also developed by Golosov and Reichman for two-level spin--boson models. We remind the readers that intra-electron correlation is distinguished from the static and dynamical correlation in quantum electronic structure, or the correlation between nuclei and electron\cite{lang2021generalized}. We 
generalize the concept of intra-electron correlation to both fully linearized mapping approaches and partially linearized mapping approaches. We prove that the correct intra-electron correlation can improve the short-time accuracy in various chemical motivated models. We also examine the intra-electron correlations for several popular mapping approaches. For each approach, we give either a proof that it correctly samples the intra-electron correlation or provide an explicit violation example. The mapping approaches considered in this article are Ehrenfest\cite{domcke2004conical}, LSC-IVR\cite{sun1998semiclassical}, PBME\cite{kim2008quantum,kelly2012mapping,nassimi2010analysis}, four different tracelss MMST\cite{saller2019identity,saller2019improved} (also named as modified LSC (mLSC)), spin linearized semiclassical method (Spin-LSC) with both full sampling and focus sampling\cite{runeson2019spin,runeson2020generalized}, partially linearized density matrix method\cite{huo2011communication,huo2013communication} (PLDM), and spin partially linearized denstiy matrix method\cite{mannouch2020partially,mannouch2020partially2} (Spin-PLDM) with both full sampling and focus sampling. Our theoretical analysis provides a measure for the short-time accuracy of mapping approaches.

This work is organized as follows. In Sec.~II, we propose the general expression for the electronic expectation value, and we establish the rigorous connection between intra-electron correlation and short-time accuracy for the methods considered in this article. In Sec.~III, we examine the intra-electron correlation for several popular mapping approaches. We either give a proof or an explicit violation example for each of the methods. In Sec.~IV, we further analyze the short-time accuracy for chemical motivated models. We also use the analysis to explain the accuracy of previous numerical benchmarks including Tully models, spin-boson models, and cavity-modified molecular dynamics. In Sec.~\ref{sec:conclusions}, we summarize the results. The proof that traceless MMST and Spin-LSC capture the intra-electron correlation correctly for two-level systems is listed in the Appendix A. The explicit time derivatives expressions are listed in the Appendix B.

\section{Theory}
This section contains four subsections. In the first subsection, we briefly review the general framework of mapping approaches. The force expression and evaluation of observables of fully linearized methods and partially linearized methods are introduced in the second and third subsection, respectively. In the last subsection, we re-write the methods in a unified expression, then establish the rigorous connection between intra-electron correlation and short-time accuracy. 

\subsection{General framework of mapping approaches}

Consider a non-adiabatic Hamiltonian in the diabatic representation of nuclei in one spatial dimension coupled to $S$ electronic states, 
\begin{align}
    \label{eq:Ham}
    \hat{H} & = \frac{\hat{p}^2}{2m} + U(\hat{x}) + \hat{V}(\hat{x})\,,\\
           \hat{V}(\hat{x}) & = \sum_{kl}^{S}|k\rangle V_{kl}(\hat{x})\langle l|\,,
\end{align}
where $m$ is the mass of the particle, and $\hat{x}$ and $\hat{p}$ are the nuclear position and momentum operators, respectively. For convenience, we prepare the initial product states in the form $\rho(0) = \rho_{\rm nuc}(0)\bigotimes\rho_{\rm el}(0)$ and $\rho_{\rm el}(0) = \ket{r}\bra{r}$. For the explicit examples that demonstrate a wrong intra-electron correlation sampling, we set $r=1$. We point out that the methods considered in this article, some of which have correct intra-electron correlation for initial diagonal states, also have correct intra-electron correlation for non-diagonal states, and even arbitrary electronic operators. In comparison, GDTWA can have correct intra-electron correlation for diagonal states, while correct intra-electron correlation for non-diagonal initial states cannot be guaranteed.

The mapping approaches considered in this article belong to two classes, fully linearized and partially linearized methods. Both methods suppose that the quantum dynamics can be approximated as the classical dynamics in the classical phase space. 

The treatment of the nuclear degrees of freedom is rather straightforward. The initial nuclei phase space distribution is selected as the Wigner function of the nuclear density matrix (up to a prefactor)\cite{wigner1932phys,hillery1984distribution,polkovnikov2010phase},
\begin{align}\label{eq:Wigner}
    W_{\rm nuc}(x_0,p_0) &=\frac{1}{2\pi}(\rho_{\rm nuc}(0))_W(x_0,p_0)\,, \nonumber \\
    (\rho_{\rm nuc}(0))_W(x_0,p_0) &:= \int d\eta \bra{x-\frac{\eta}{2}}\rho_{\rm nuc}(0)\ket{x+\frac{\eta}{2}}e^{ip\eta}\,.
\end{align}
The generalization to multidimensional systems is straightforward by replacing $x,p,\eta$ by the corresponding vectors and the pre-factor $(2\pi)^{-1}$ by $(2\pi)^{-d}$, where $d$ is the dimension of the system. Specifically, for any scalar function $g$, the Wigner transformation gives
\begin{align}\label{eq:WignerExplicit}
    g(x) &= (g(\hat{x}))_W(x,p)\,, \nonumber \\
    g(p) &= (g(\hat{p}))_W(x,p)\,, \nonumber \\
    pg(x) &= (\frac{\hat{p}g(\hat{x}) + g(\hat{x})\hat{p}}{2})_W(x,p)\,, \nonumber \\
    p^2g(x) &= (\frac{\hat{p}^2g(\hat{x}) + g(\hat{x})\hat{p}^2 + 2\hat{p}g(\hat{x})\hat{p}}{4})_W(x,p)
\end{align}

Consider an arbitrary nuclei observable $\hat{O}_{\rm nuc}$. The statistical average of its Wigner function $O_{w,\rm nuc}(x_0,p_0):=(\hat{O}_{\rm nuc})_W(x_0,p_0)$ over the Wigner distribution is exactly identical to its quantum expectation value $\braket{\hat{O}_{\rm nuc}}$
\begin{align}\label{eq:WignerIdentity}
    &\braket{\hat{O}_{\rm nuc}} = \Tr{\rho(0)\hat{O}_{\rm nuc}}_{\rm q} \nonumber \\
    &= \frac{1}{2\pi}\int dx_0dp_0 (\rho_{\rm nuc}(0))_W(x_0,p_0)(\hat{O}_{\rm nuc})_W(x_0,p_0) \nonumber \\
    &= \int dx_0dp_0 W_{\rm nuc}(x_0,p_0)O_{w,\rm nuc}(x_0,p_0)\,,
\end{align}
where $\Tr{}_{\rm q}$ represents the trace over all quantum DoFs. In this article, we use operator with subscript ``nuc" to represent the nuclear observables. Further below, we will denote traceless electronic observables as $\hat{O}$, and arbitrary electronic observables as $\hat{B}$ and $\hat{C}$.

Unlike the nuclei DoF, the electronic subsystem is discrete. A more severe problem is that the electronic DoF does not have an immediate expression in position and momentum operators. Various different mapping approaches are developed to resolve this problem. The basic idea of those methods are similar, i.e., to find the proper phase space to describe the electronic subsystem. Successful attempts include the $SU(S)$ Schwinger boson (MMST harmonic oscillator) phase space\cite{meyera1979classical,cotton2013symmetrical,stock1997semiclassical,cotton2013symmetrical2,miller2017classical,saller2019identity,saller2019improved,sun1998semiclassical,kim2008quantum,kelly2012mapping,huo2011communication,huo2013communication,hsieh2012nonadiabatic,hsieh2013analysis,kapral1999mixed,stock1999flow,muller1999flow,huo2012consistent}, Stratonovich--Weyl (SW) spin phase space\cite{stratonovich1957statistical,runeson2019spin,runeson2020generalized,mannouch2020partially,mannouch2020partially2},  Wootters' spin discrete phase space\cite{wootters1987wigner,zhu2019generalized,lang2021generalized}, action-angle phase space\cite{cotton2013symmetrical,cotton2013symmetrical2}, etc. In this article, we mainly focus on $SU(S)$ Schwinger boson (harmonic oscillator) phase space and Stratonovich--Weyl spin phase space. We denote the collection of electronic phase space variables by $\Gamma$.  For all the methods considered in this article, the equations of motions (EOMs) of nuclear classical phase space variables are
\begin{equation}\label{eq:EOMG}
\begin{aligned}
    \dot{x}_t & = p_t/m\,, \\
    \dot{p}_t & = -\partial_{x_t}U(x_t) + F(\Gamma_t,x_t)\,, \\
\end{aligned}
\end{equation}
where $F(\Gamma_t,x_t)$, as a $\Gamma_t,x_t$-dependent function, is the electron back-action force to the nuclei, and it can be expressed as
\begin{equation}\label{eq:defF}
F(\Gamma_t,x_t) = -\Tr{\frac{\partial\hat{V}(x_t)}{\partial x_t}\hat{F}(\Gamma_t)}\,,
\end{equation}
where, depending on the situation, $\Tr{}$ represents the trace over the electronic DoFs or the usual matrix trace. The EOMs of $\hat{F}(\Gamma_t)$ have a unified expression,
\begin{align}\label{eq:EOMF}
    \frac{d}{dt}\hat{F}(\Gamma_t) &= i[\hat{F}(\Gamma_t),\hat{V}(x_t)]\,,
\end{align}
The explicit form of $\hat{F}(\Gamma_t)$, and the way to evaluate observables will be given in the following two subsections.

\subsection{Fully linearized methods}

For the fully linearized methods considered in this article, the electronic phase space variables are the positions and momenta of $S$ fictitious particles\cite{sun1998semiclassical,kim2008quantum,runeson2019spin,runeson2020generalized,saller2019identity,saller2019improved}, $\Gamma_t = (X_t,P_t) = (X_1(t),X_2(t),\cdots,X_S(t),P_1(t),P_2(t),\cdots,P_S(t))$. In these methods, the treatment of the electronic DoFs is similar in spirit to the truncated Wigner approximation (TWA) for pure nuclei DoFs\cite{polkovnikov2010phase}. The initial electronic phase space distribution, $W_{\rm el}(\Gamma_0)$, is generated according to $\rho_{\rm el}(0)$, then the phase space variables are sampled according to the initial phase space distribution and propagated. The expectation of any traceless electronic observable $\hat{O}$ is evaluated as the statistical average of the corresponding phase space expression over the phase space trajectories $\braket{\hat{O}(t)}_{\rm m}$. For convenience, we define the matrix $\hat{K}(X_t,P_t)$
\begin{align}
    K_{mn}(X_t,P_t) &= \frac{(X_m(t)+iP_m(t))(X_n(t)-iP_n(t))}{2}\,.
\end{align}
Then, the explicit forms of $\hat{F}(\Gamma_t)$, and the EOMs of $\Gamma_t$ are
\begin{align}
    \hat{F}_{\rm f}(\Gamma_t) &= \hat{K}(X_t,P_t) \\
    \dot{X}_m(t) &= \sum_n \frac{V_{mn}(x_t) + V_{nm}(x_t)}{2}P_n(t) \nonumber \\
    &- i\sum_n \frac{V_{mn}(x_t) - V_{nm}(x_t)}{2}X_n(t) \label{eq:EOMX}\\
    \dot{P}_m(t) &= -\sum_n \frac{V_{mn}(x_t) + V_{nm}(x_t)}{2}X_n(t) \nonumber \\
    &- i\sum_n \frac{V_{mn}(x_t) - V_{nm}(x_t)}{2}P_n(t) \label{eq:EOMP}\,.
\end{align}
The corresponding phase space expression of $\hat{O}$ is given by the replacement
\begin{equation}
    \hat{O} \rightarrow\Tr{\hat{O}\hat{K}(X,P)}\,,
\end{equation}
and the time dependent expectation value $\braket{\hat{O}(t)}$ is approximated as
\begin{align}
    &\braket{\hat{O}(t)} \approx \braket{\hat{O}(t)}_{\rm m} \nonumber \\
    &= \int dx_0dp_0d\Gamma_0 W_{\rm nuc}(x_0,p_0)W_{\rm el}(\Gamma_0)\Tr{\hat{O}\hat{K}(X_t,P_t)}\,.
\end{align}
Different fully linearized methods choose different initial electronic phase space distributions $W_{\rm el}(\Gamma_0)$. The explicit expression of $W_{\rm el}(\Gamma_0)$ for each method will be listed in the corresponding section.

\subsection{Partially linearized methods}

In the partially linearized methods, the initial electronic phase space sampling is achieved by inserting the ``closure relation" of either coherent state\cite{hsieh2012nonadiabatic,huo2011communication} (Eq.~(\ref{eq:csIdentity}) for PLDM) or spin coherent state\cite{mannouch2020partially,mannouch2020partially2} (Eq.~(\ref{eq:scsfullIdentiy},\ref{eq:scsfocIdentity}) for Spin-PLDM) between $\rho_{\rm el}(0)$ and forward/backward propagator $e^{-i\hat{H}t}/e^{i\hat{H}t}$. Therefore, two sets of electronic phase space variables $\hat{\Gamma}=(X,P,X^\prime,P^\prime)$ are required. We suppose that $(X_t,P_t)$ is used to represent the forward propagator and $(X_t^\prime,P_t^\prime)$ is used to represent the backward propagator.
The forward and backward propagator in each single phase space trajectory are approximated as $\hat{W}(X_0,P_0,t)$ and $\hat{W}^\dagger(X_0^\prime,P_0^\prime,t)$, and they satisfy the following initial conditions and EOMs,
\begin{align}\label{eq:Wmatrix}
    \hat{W}(X_0,P_0,0) &= \hat{K}(X_0,P_0) - \frac{\gamma\hat{I}}{2}\,, \nonumber \\
    \frac{d}{dt}\hat{W}(X_0,P_0,t) &= -i\hat{V}(x_t)\hat{W}(X_0,P_0,t)\,, 
\end{align}
and analogously for $\hat{W}^\dagger(X_0^\prime,P_0^\prime,t)$ by taking the hermitian conjugate and replacing $X_0,P_0$ by $X_0^\prime,P_0^\prime$. Here, $\gamma$ is the zero-point energy (ZPE) parameter and differs in PLDM and Spin-PLDM. Furthermore, each set of electronic phase space variables has the same form of EOMs as the fully linearized scenario, i.e., Eqs.~(\ref{eq:EOMX}) and~(\ref{eq:EOMP}).

The nuclear DoF moves along the mean force of forward and backward electronic phase space variables, which leads to
\begin{align}
    \hat{F}_{\rm p}(\Gamma_t)  = \frac{\hat{K}(X_t,P_t) + \hat{K}(X_t^\prime,P_t^\prime)}{2}\,,
\end{align}
up to an arbitrary additional matrix which is proportional to the identity due to the traceless of $\hat{V}$. The expectation value of a traceless operator $\hat{O}$ reads as
\begin{equation}\label{eq:ObsPL}
\begin{aligned}
    &\braket{\hat{O}(t)}_{\rm m} = \int dx_0dp_0d\Gamma_0 W_{\rm nuc}(x_0,p_0)h(X_0,P_0) \\
    &\times h(X_0^\prime,P_0^\prime)\Tr{\hat{W}(X_0,P_0,t)\rho_{\rm el}(0)\hat{W}^\dagger(X_0^\prime,P_0^\prime,t)\hat{O}}
    \,,
\end{aligned}
\end{equation}
where $h(X_0,P_0)$ and $h(X_0^\prime,P_0^\prime)$ are method-dependent sampling weights that appear in the ``closure relations." For details, see Secs.~III. Fully linearized approaches can be regarded as approximations of partially linearized approaches \cite{mannouch2020partially,mannouch2020partially2,mannouch2022partially}. Usually, the partially linearized approaches provide more accurate results than the fully linearized approaches.

\subsection{Intra-electron correlation and short-time accuracy}
In the previous subsections, we have briefly reviewed the approaches considered in this article. Now we are at the position to re-write them in a unified expression. With the help of such a re-formulation, we will establish the rigorous connection between short-time accuracy and intra-electron correlation.

For the methods considered in this article, the expectation value $\braket{\hat{O}(t)}_{\rm m}$ can be expressed through the following unified form
\begin{equation}\label{eq:ObsGene}
\begin{aligned}
    &\braket{\hat{O}(t)}_{\rm m} = \int dx_0dp_0W_{\rm nuc}(x_0,p_0) \llangle\Tr{\hat{A}(\Gamma_0,t)\hat{O}}\rrangle \\
    &:= \int dx_0dp_0d\Gamma_0 W_{\rm nuc}(x_0,p_0)f(\Gamma_0)\Tr{\hat{A}(\Gamma_0,t)\hat{O}}
    \,,
\end{aligned}
\end{equation}
where $\hat{A}(\Gamma_0,t)$ and $f(\Gamma_0)$ are method-dependent and initial state-dependent quantities, and the double angle bracket means the integration over the weight factor $f(\Gamma_0)$. The EOM of $\hat{A}(\Gamma_0,t)$ is unified
\begin{equation}\label{eq:EOMAGamma}
    \frac{d}{dt}\hat{A}(\Gamma_0,t) = i[\hat{A}(\Gamma_0,t),\hat{V}(x_t)]\,.
\end{equation}
To summarize, the EOMs Eq.~(\ref{eq:EOMG},\ref{eq:defF},\ref{eq:EOMF},\ref{eq:EOMAGamma}), and the observables evaulation expression Eq.~(\ref{eq:ObsGene}) give the unnified expressions of mapping approaches considered in this article. For the fully linearized approaches,
\begin{align}
    f(\Gamma_0) &= W_{\rm el}(X_0,P_0)\,, \\
    \hat{A}(\Gamma_0,0) &= \hat{F}_{\rm f}(\Gamma_0)= \hat{K}(X_0,P_0) \,,
\end{align}
while for the partially linearized approaches,
\begin{align}
    f(\Gamma_0) &= h(X_0,P_0)h(X_0^\prime,P_0^\prime)\,, \\
    \hat{A}(\Gamma_0,0) &= \hat{W}(X_0,P_0,0)\rho_{\rm el}(0)\hat{W}^\dagger(X_0^\prime,P_0^\prime,0)\,, \\
    \hat{F}_{\rm p}(\Gamma_0) &= \frac{\hat{W}(X_0,P_0,0) + \hat{W}^\dagger(X_0^\prime,P_0^\prime,0)}{2}\,.
\end{align}

With the help of this unified form, it becomes apparent that the first as well as second time derivatives of $\braket{\hat{O}(t)}_{\rm m}$ and $\braket{\hat{O}(t)}$ at $t =0$ coincide. 
The difference between the third time derivatives at $t = 0$ is (the individual expressions are reported in Appendix B)
\begin{align}\label{eq:bare}
    &\frac{d^3}{dt^3}\braket{\hat{O}(t)}_{\rm m}|_{t=0} - \frac{d^3}{dt^3}\braket{\hat{O}(t)}|_{t=0} =  \nonumber\\
    &-i\frac{1}{2m}\braket{[\hat{O},\frac{\partial\hat{V}(\hat{x}_0)}{\partial \hat{x}_0}]\frac{\partial\hat{V}(\hat{x}_0)}{\partial \hat{x}_0} +  \frac{\partial\hat{V}(\hat{x}_0)}{\partial \hat{x}_0}[\hat{O},\frac{\partial\hat{V}(\hat{x}_0)}{\partial \hat{x}_0}]} \nonumber \\ 
    &+i\int dx_0dp_0W_{\rm nuc}(x_0,p_0)\nonumber\\
     &\times\llangle\Tr{\hat{A}(\Gamma_0,0)[\hat{O}, \frac{\partial\hat{V}(x_0)}{\partial x_0}]}\Tr{\frac{\partial\hat{V}(x_0)}{\partial x_0}\hat{F}(\Gamma_0)}\frac{1}{m}\rrangle     \,.
\end{align}
After integrating out the nuclear DoF, the vanishing of Eq.~(\ref{eq:bare}) requires 
\begin{align}\label{eq:IntraEleDef}
    &\Tr{\rho_{\rm el}(0)\frac{\hat{O}_1\hat{O}_2 + \hat{O}_2\hat{O}_1}{2}} \nonumber \\&= \llangle \Tr{\hat{A}({\Gamma_0,0})\hat{O}_1}\Tr{\hat{F}(\Gamma_0)\hat{O}_2}\rrangle\,,
\end{align}
where $\hat{O}_1$ and $\hat{O}_2$ are two arbitrary traceless electron operators, or equivalently, they are arbitrary generalized Gell-Mann matrices. Thus, Eq.~({\ref{eq:IntraEleDef}}) is the definition of the methods with correct intra-electron correlation sampling, which generically have a higher short-time accuracy than the methods with wrong intra-electron correlations. We stress that the $\mathcal{O}(t^3)/\mathcal{O}(t^2)$ accuracy for the methods with/without correct intra-electron correlations can be improved when the Hamiltonian $\hat{H}$ and evaluation observable $\hat{O}$ have specific forms. Detailed discussions can be found in the Sec.~IV.

The intra-electron correlation can be written more explicitly, for the fully linearized methods,
\begin{align}\label{eq:IntraFulDef}
    &\Tr{\rho_{\rm el}(0)\frac{\hat{O}_1\hat{O}_2 + \hat{O}_2\hat{O}_1}{2}} \nonumber\\
    &= \llangle \Tr{\hat{K}(X_0,P_0)\hat{O}_1}\Tr{\hat{K}(X_0,P_0)\hat{O}_2}\rrangle\,,
\end{align}
and for the partially linearized methods,
\begin{align}\label{eq:IntraEParDef}
    &\Tr{\rho_{\rm el}(0)\frac{\hat{O}_1\hat{O}_2 + \hat{O}_2\hat{O}_1}{2}} \nonumber\\
    &= \llangle \Tr{\hat{W}(X_0,P_0,0)\rho_{\rm el}(0)\hat{W}^\dagger(X_0^\prime,P_0^\prime,0)\hat{O}_1} \nonumber\\
    &\times\Tr{\frac{\hat{K}(X_0,P_0) + \hat{K}(X_0^\prime,P_0^\prime)}{2}\hat{O}_2}\rrangle\,.
\end{align}
The intra-electron correlation for the fully linearized methods considered in this article is identical to its original definition\cite{zhu2019generalized,lang2021generalized}. However, modifications are required for the partially linearized methods.  Interestingly, the definition of intra-electron correlation for the partially linearized methods looks asymmetric. In fact, such asymmetry represents the different roles of $\hat{O}_1$ and $\hat{O}_2$ in the partially linearized methods. The term $\Tr{\hat{W}(X_0,P_0,0)\rho_{\rm el}(0)\hat{W}^\dagger(X_0^\prime,P_0^\prime,0)\hat{O}_1}$ represents the observable evaluations and $\Tr{\frac{\hat{K}(X_0,P_0) + \hat{K}(X_0^\prime,P_0^\prime)}{2}\hat{O}_2}$ represents the back-action force. As a comparison, the back-action force representations and the observable evaluations for the fully linearized methods considered in this article are symmetrical.

\section{Classification}
In this section, we examine the intra-electron correlation for several popular methods. Those methods can be classified into three categories, namely, wrong intra-electron correlation, correct intra-electron correlation only for 2-level systems, and correct intra-electron correlation. Three traditional methods, Ehrenfest, PBME, and LSC-IVR fall into the first category. mLSC$/\phi^1\phi^2$, mLSC$/\phi^2\phi^2$, full spin-LSC, and focus spin-LSC fall into the second category. mLSC$/\phi^1\phi^1$, mLSC$/\phi^2\phi^1$, single Wigner mLSC, full spin-PLDM, focus spin-PLDM, and PLDM fall into the third category. Readers who are not interested in the proof of the respective classifications or explicit violation examples can skip this section.

\subsection{Ehrenfest, PBME and LSC-IVR}
In this subsection, we examine the intra-electron correlation Eq.~({\ref{eq:IntraFulDef}}) for MMST mappings. The Ehrenfest method is the simplest MMST mapping approach, in which the electronic DoFs are sampled on a single point\cite{domcke2004conical}
\begin{align}
    f_{\rm Ehrenfest}(\Gamma) &= \frac{1}{\pi^S}\delta(X_r^2 + P_r^2 - 2)\prod_{n\neq r}\delta(X_n^2 + P_n^2) \,.
\end{align}

Other, more advanced MMST approaches\cite{sun1998semiclassical,kim2008quantum,saller2019identity,saller2019improved} map the electronic system onto $S$ harmonic oscillators within the singly-excited harmonic oscillators (SEO) subspace ($SU(S)$ Schwinger bosons). The electronic state $\ket{m}$ maps onto $\ket{M}$ which represents the $m$-th oscillator being on the first excited state and other oscillators being on the ground states. The projector on the SEO subspace is defined as $\hat{\Pi}_1 = \sum_M \ket{M}\bra{M}$. The creation, annihilation, position, and momentum operators for the $m$-th oscillator are defined as $\hat{a}_m^{\dagger}$, $\hat{a}_m$, $\hat{X}_m = (\hat{a}_m^{\dagger} + \hat{a}_m)/\sqrt{2}$, and $\hat{P}_m=i(\hat{a}_m^{\dagger} - \hat{a}_m)/\sqrt{2}$, respectively.

The mapping of $\ket{m}\bra{n}$ can take two forms: it can either be a pure creation-annihilation term\cite{kim2008quantum} $\hat{a}_m^{\dagger}\hat{a}_n$ or a creation-annihilation term projected onto the SEO subspace $\ket{M}\bra{N}=\hat{a}_m^{\dagger}\hat{a}_n\hat{\Pi}_1 =\hat{a}_m^{\dagger}\hat{\Pi}_0\hat{a}_n  $\cite{sun1998semiclassical}, where $\hat{\Pi}_0$ is the projector of the ground state of fictitious oscillators. The trace of the product of electronic DoFs can be represented as the trace over all quantum DoFs of the mapping harmonic oscillators,
\begin{align}\label{eq:MappingHO}
    &\Tr{\hat{B}\hat{C}} = \Tr{ (\sum_{m,n}B_{mn}\hat{a}^\dagger_m\hat{a}_n)(\sum_{k,l}C_{kl}\hat{a}^\dagger_k\hat{a}_l\hat{\Pi}_1) }_{\rm q} \nonumber \\
    &= \Tr{ (\sum_{m,n}B_{mn}\hat{a}^\dagger_m\hat{a}_n\hat{\Pi}_1)(\sum_{k,l}C_{kl}\hat{a}^\dagger_k\hat{a}_l\hat{\Pi}_1) }_{\rm q} \,,
\end{align}
where the operator $\hat{B}$ is replaced by either $(\sum_{m,n}B_{mn}\hat{a}^\dagger_m\hat{a}_n)$ or $(\sum_{m,n}B_{mn}\hat{a}^\dagger_m\hat{a}_n\hat{\Pi}_1)$, and the SEO projector $\hat{\Pi}_1$ should appear at least once (analogously, one could drop $\hat{\Pi}_1$ from the sum over $k,l$ in the second line). The expression for the trace of the product of multiple operators generalization is similar. With the help of Eq.~(\ref{eq:MappingHO}), one can use the standard Wigner transformation Eq.~(\ref{eq:WignerIdentity}) to express the trace of the product of electronic operators. We list several useful Wigner transformations for the mapping systems,
\begin{align}
    (\hat{a}_m^{\dagger}\hat{a}_n)_W(X,P) &= \frac{(X_m-iP_m)(X_n+iP_n) - \delta_{mn}}{2}\,, \nonumber \\
    (\hat{a}_m^{\dagger}\hat{a}_n\hat{\Pi}_1)_W(X,P) &= \frac{(X_m-iP_m)(X_n+iP_n) - \delta_{mn}/2}{2}\phi\,,
\end{align}
\begin{align}\label{eq:WImLSC}
    &(\sum_{m,n}O_{mn}\hat{a}^\dagger_m\hat{a}_n)_W(X_0,P_0) = \Tr{\hat{O}\hat{K}(X_0,P_0)} \,, \nonumber \\
    &(\sum_{m,n}O_{mn}\hat{a}^\dagger_m\hat{a}_n\hat{\Pi}_1)_W(X_0,P_0) = \Tr{\hat{O}\hat{K}(X_0,P_0)}\phi \,,
\end{align}
\begin{align}\label{eq:WQrmLSC1}
    &(\sum_{m,n,k,l}\frac{O_{1,mn}O_{2,kl}(\hat{a}^\dagger_m\hat{a}_n\hat{a}^\dagger_k\hat{a}_l + \hat{a}^\dagger_k\hat{a}_l\hat{a}^\dagger_m\hat{a}_n) }{2} )_W(X_0,P_0) \nonumber \\
    &= \Tr{\hat{O}_1\hat{K}(X_0,P_0)}\Tr{\hat{O}_2\hat{K}(X_0,P_0)} - \frac{\Tr{\hat{O}_1\hat{O}_2}}{4}\,,
\end{align}
\begin{align}\label{eq:GroundProjectormLSC}
    ( \hat{\Pi}_0 )_W(X_0,P_0) &= \frac{\phi}{4}\,, \nonumber \\
     \phi(\sum_r \hat{a}^\dagger_r\hat{a}_r)_W(X_0,P_0)  &= ( \hat{\Pi}_1 )_W(X_0,P_0) - \frac{S}{4}\phi\,.
\end{align}
where
\begin{equation}
    \phi = 2^{(S+2)}\exp{-\sum_{m=1}^S(X_m^2+P_m^2)}\,,
\end{equation} 
is a conserved quantity.

Both PBME\cite{kim2008quantum} and LSC-IVR\cite{sun1998semiclassical} use $(\hat{a}_r^{\dagger}\hat{a}_r\hat{\Pi}_1)_W(X_0,P_0)$ as the initial electronic phase space. The major difference between the two methods is in the mapping procedure of observables. PBME maps the electronic observable onto pure creation--annihilation terms, which yields (for a traceless observable $\hat{O}$)
\begin{align}
    f_{\rm PBME}(\Gamma) &= \frac{1}{2(2\pi)^S}(X_r^2 + P_r^2 - \frac{1}{2})\phi\,,
\end{align}
while LSC-IVR maps the observables onto a creation--annihilation term projected onto the SEO subspace, which yields an effective sampling (for a traceless observable $\hat{O}$)
\begin{align}
    f_{\rm LSC-IVR}(\Gamma) &= \frac{1}{2(2\pi)^S}(X_r^2 + P_r^2 - \frac{1}{2})\phi^2\,.
\end{align}

We point out that neither of Ehrenfest nor PBME or LSC-IVR is able to correctly sample the intra-electron correlation, even for a two-level system. As an explicit violation example, consider $\hat{O}_1 = \hat{O}_2 = \ket{1}\bra{2} + \ket{2}\bra{1}$. Then,  
\begin{equation}
    \braket{\frac{\hat{O}_1\hat{O}_2 + \hat{O}_2\hat{O}_1}{2}} = 1 \,,
\end{equation}
while 
\begin{align}
    \llangle \Tr{\hat{K}(X_0,P_0)\hat{O}_1}\Tr{\hat{K}(X_0,P_0)\hat{O}_2}\rrangle_{\rm Ehrenfest} &= 0\,,\nonumber\\ 
    \llangle \Tr{\hat{K}(X_0,P_0)\hat{O}_1}\Tr{\hat{K}(X_0,P_0)\hat{O}_2}\rrangle_{\rm LSC-IVR} &= \frac{1}{2}\,,\nonumber\\ 
    \llangle \Tr{\hat{K}(X_0,P_0)\hat{O}_1}\Tr{\hat{K}(X_0,P_0)\hat{O}_2}\rrangle_{\rm PBME} &= \frac{3}{2}\,.
\end{align}

\subsection{mLSC}
In recent years, modified LSC methods\cite{saller2019identity,saller2019improved} have been developed to improve the accuracy of MMST mappings. The basic idea of mLSC methods is to separate the initial state into identity and traceless parts,
\begin{align}\label{eq:separation}
    \ket{r}\bra{r} &= \frac{1}{S}(\hat{I} + \hat{Q}_r)\,, \\
    \hat{Q}_r &= S\ket{r}\bra{r} - \hat{I} = (S-1)\ket{r}\bra{r} - \sum_{m \neq r}^S \ket{m}\bra{m}\,,
\end{align}
and calculate their contribution to the expectation value separately, 
\begin{widetext}
\begin{align}
    \Tr{\rho_{\rm nuc}\otimes\hat{I}e^{i\hat{H}t}\hat{O}e^{-i\hat{H}t}}_{\rm q} &\approx \frac{1}{(2\pi)^S}\int dx_0dp_0d\Gamma_0 W_{\rm nuc}(x_0,p_0)\phi^a\Tr{\hat{K}(X_t,P_t)\hat{O}}\,, \\
    \Tr{\rho_{\rm nuc}\otimes\hat{Q}_re^{i\hat{H}t}\hat{O}e^{-i\hat{H}t}}_{\rm q} &\approx \frac{1}{(2\pi)^S}\int dx_0dp_0d\Gamma_0 W_{\rm nuc}(x_0,p_0)\phi^b \Tr{\hat{K}(X_0,P_0)\hat{Q}_r}\Tr{\hat{K}(X_t,P_t)\hat{O}} \,,
\end{align}

\end{widetext}
where $a,b = 1,2$. The different $a,b$ combinations give different mLSC/$\phi^a\phi^b$ methods\cite{gao2020benchmarking}. For completeness, we list the explicit form of $f(\Gamma)$ for mLSC/$\phi^a\phi^b$,
\begin{align}
    f_{\rm mLSC/\phi^a\phi^b}(\Gamma) &= \frac{1}{S(2\pi)^S}\{ \phi^a + \frac{\phi^b}{2}[S(X_r^2 + P_r^2) \nonumber \\
    &\qquad\qquad - \sum_{m}^S(X_m^2+P_m^2) ]\}\,.
\end{align}

The $\phi^1$ sampling for the identity operator can be obtained by mapping the initial $\hat{I}$ and observable $\hat{O}(t)$ onto $(1)_W(X_0,P_0)$ and $(\sum_{mn}O_{mn}\hat{a}_m^{\dagger}\hat{a}_n\hat{\Pi}_1)_W(X_t,P_t)$, while the prescription of $\phi^2$ sampling for the identity operator is \textit{ad hoc}. For the tracless part $\hat{Q}_r$, both $\phi^1$ and $\phi^2$ sampling map $\hat{Q}_r$ onto $(\sum_{mn}{Q}_{r,mn}\hat{a}_m^\dagger\hat{a}_n\hat{\Pi}_1)_W(X_0,P_0)$, while they use different mapping for observables $\hat{O}(t)$, i.e., $(\sum_{mn}O_{mn}\hat{a}_m^{\dagger}\hat{a}_n)_W(X_t,P_t)$ for $\phi^1$ and $(\sum_{mn}O_{mn}\hat{a}_m^{\dagger}\hat{a}_n\hat{\Pi}_1)_W(X_t,P_t)$ for $\phi^2$. One can also map $\hat{Q}_r$ onto $(\sum_{mn}{Q}_{r,mn}\hat{a}_m^\dagger\hat{a}_n)_W(X_0,P_0)$ and map $\hat{O}(t)$ onto $(\sum_{mn}O_{mn}\hat{a}_m^{\dagger}\hat{a}_n\hat{\Pi}_1)_W(X_t,P_t)$ to obtain the $\phi^1$ sampling of the traceless part.

There is also a mLSC named single Wigner mLSC\cite{saller2019identity}, which uses the mapping $(\hat{a}_r^{\dagger}\hat{a}_r)_W(X_0,P_0)$ for the initial operator and $(\sum_{mn}O_{mn}\hat{a}_m^{\dagger}\hat{a}_n\hat{\Pi}_1)_W(X_t,P_t)$ for observables (for a traceless $\hat{O}$), which yields
\begin{align}
    f_{\rm SingleWigner}(\Gamma) &= \frac{1}{2(2\pi)^S}(X_r^2 + P_r^2 - 1)\phi\,.
\end{align}
One can also use the separation Eq.~(\ref{eq:separation}) in single Wigner mLSC. The traceless part is equivalent to the $\phi^1$ mapping, and the identity part is obtained by mapping  $\hat{I}$ and $\hat{O}(t)$ onto $(\sum_m \hat{a}_m^{\dagger}\hat{a}_m)_W(X_0,P_0)$ and $(\sum_{mn}O_{mn}\hat{a}_m^{\dagger}\hat{a}_n\hat{\Pi}_1)_W(X_t,P_t)$, respectively.

For mLSC methods, it is insightful to examine the intra-electron correlation for the identity operator and traceless operator separately. For the identity operator, both $a = 1,2$ samplings hold the intra-electron correlation
\begin{align} \label{eq:mLSCIdIntra}
    \Tr{\hat{O}_1\hat{O}_2} &= \frac{1}{(2\pi)^S}\int d\Gamma_0 \phi^a\Tr{\hat{K}(X_0,P_0)\hat{O}_1} \nonumber \\
    &\qquad \qquad \qquad \quad\,\, \times\Tr{\hat{K}(X_0,P_0)\hat{O}_2}\,.
\end{align}
Applying Eq.~(\ref{eq:WignerIdentity}) to Eq.~(\ref{eq:MappingHO}) and using the Wigner transformation Eqs.~(\ref{eq:WImLSC}), Eq.~(\ref{eq:mLSCIdIntra}) can be obtained immediately.

The identity part of single Wigner mLSC also has the correct intra-electron correlation, 
\begin{align} \label{eq:mLSCIdIntraSingleWigner}
    \Tr{\hat{O}_1\hat{O}_2} &= \frac{1}{(2\pi)^S}\int d\Gamma_0 \phi(\sum_r \hat{a}^\dagger_r\hat{a}_r)_W(X_0,P_0)\nonumber \\
    &\times \Tr{\hat{K}(X_0,P_0)\hat{O}_1}\Tr{\hat{K}(X_0,P_0)\hat{O}_2}\,.
\end{align}
With the help of Wigner transformations Eq.~(\ref{eq:WQrmLSC1}) and ~(\ref{eq:GroundProjectormLSC}), and applying Eq.~(\ref{eq:WignerIdentity}) to the fictitious oscillators, the right hand side of Eq.~(\ref{eq:mLSCIdIntraSingleWigner}) can be expressed as
\begin{align}
    &\frac{1}{(2\pi)^S}\int d\Gamma_0  [( \hat{\Pi}_1 )_W(X_0,P_0) - S( \hat{\Pi}_0 )_W(X_0,P_0)]\nonumber \\
    &\times [(\sum_{m,n,k,l}\frac{O_{1,mn}O_{2,kl}\hat{a}^\dagger_m\hat{a}_n\hat{a}^\dagger_k\hat{a}_l + \hat{a}^\dagger_k\hat{a}_l\hat{a}^\dagger_m\hat{a}_n) }{2} )_W(X_0,P_0) \nonumber \\
    &+ \frac{\Tr{\hat{O}_1\hat{O}_2}}{4}] = \Tr{\hat{O}_1\hat{O}_2} + S\frac{\Tr{\hat{O}_1\hat{O}_2}}{4} - S\frac{\Tr{\hat{O}_1\hat{O}_2}}{4}\,,
\end{align}
which equals to the left hand side of the equation.

For the traceless operator, only the approach with $b = 1$ can faithfully sample the intra-electron correlation for systems with arbitrary $S$,
\begin{align}\label{eq:mLSCQdIntra}
    &\Tr{\hat{Q}_r\frac{\hat{O}_1\hat{O}_2 + \hat{O}_2\hat{O}_1}{2}} =\frac{1}{(2\pi)^S}\int d\Gamma_0 \phi \nonumber \\
    &\Tr{\hat{K}(X_0,P_0)\hat{Q}_r}\Tr{\hat{K}(X_0,P_0)\hat{O}_1}\Tr{\hat{K}(X_0,P_0)\hat{O}_2}\,.
\end{align}
The proof of Eq.~(\ref{eq:mLSCQdIntra}) is quite similar to the proof of the intra-electron correlation for the identity operator Eq.~(\ref{eq:mLSCIdIntra}). Applying Eq.~(\ref{eq:WignerIdentity}) to Eq.~(\ref{eq:MappingHO}), setting $\hat{C} = \hat{Q}_r$ and $\hat{B} = (\hat{O}_1\hat{O}_2 + \hat{O}_2\hat{O}_1)/2$, and using the Wigner transformation Eq.~(\ref{eq:WQrmLSC1}), Eq.~(\ref{eq:mLSCQdIntra}) can be obtained immediately. Apparently, the proof does not use the diagonal $\hat{Q}_r$ property, therefore, mLSC$/\phi^a\phi^1$, $a = 1,2$, and single Wigner mLSC can also capture correct intra-electron correlation for non-diagonal initial states.

We also give an explicit violation example to show how the approach with $b = 2$ fails to sample the intra-electron correlation for a traceless operator in a $S$-level system, where $S > 2$. Consider $\hat{O}_1 = \hat{O}_2 = \ket{2}\bra{3} + \ket{3}\bra{2}$. Then,  
\begin{equation}
    \Tr{\hat{Q}_r\frac{\hat{O}_1\hat{O}_2 + \hat{O}_2\hat{O}_1}{2}} = -2\,, \nonumber 
\end{equation}
while
\begin{align}
    &\frac{1}{(2\pi)^S}\int dXdP \frac{1}{2}[S(X_1^2 + P_1^2)  \nonumber \\
    &- \sum_{m}^S(X_m^2+P_m^2) ]\phi^2 (X_2X_3 + P_2P_3)^2 = -1\,.
\end{align}

In the end of this subsection, we stress that the $b = 2$ approach for traceless operator samples the intra-electron correlation perfectly for two-level system. The detailed proof is listed in the Appendix A.

\subsection{Spin-LSC}

The spin mapping based on SW representation uses the initial electronic phase space instead of the phase space of the $SU(S)$ Schwinger bosons. There are three different common SW representations\cite{runeson2019spin,runeson2020generalized}, Glauber-P representation, Husimi-Q representation, and Wigner-W representation, which correspond to different ZPE $\gamma_{\rm P} = 2$, $\gamma_{\rm Q} = 0$, and 
\begin{align}
    \gamma_{\rm W} &= \frac{R^2-2}{S} = \frac{2\sqrt{S+1} - 2}{S}\label{eq:ZPE} \,.
\end{align}
Spin-LSC based on Wigner-W representation is the most robust and accurate one\cite{runeson2019spin,runeson2020generalized}. Thus, we only consider Spin-LSC based on Wigner-W representation, and unless we specify otherwise mean the Wigner-W representation when we mention SW representation. We also use $\gamma_{\rm W}$ to represent $\gamma$ when there is no ambiguity. Interestingly, all three spin mappings can be interpreted under the framework of the constrained phase space\cite{liu2021unified,he2019new,he2021commutator,he2022new,he2021negative}, and other ZPE parameters (even negative\cite{he2021negative}) are also possible.

The SW representation in Cartesian variables for the spin system is
\begin{align}
    \Tr{\hat{B}\hat{C}} &= \int dXdP \rho_{\rm full}(X,P) \nonumber \\
    &\times\Tr{\hat{W}_{\rm scs}(X,P)\hat{B}}\Tr{\hat{W}_{\rm scs}(X,P)\hat{C}}\,,
\end{align}
where $\hat{B}$ and $\hat{C}$ are arbitrary electronic operators. The definitions of the full sampling integration weight $\rho_{\rm full}(X,P)$ and the SW transformation kernel $\hat{W}_{\rm scs}(X,P)$ are
\begin{align}
    \rho_{\rm full}(X,P) &= \frac{S!R}{\pi^S}\delta\big(\sum_{m=1}^S(X_m^2+P_m^2)-R^2\big)\,, \nonumber \\
    \hat{W}_{\rm scs}(X,P) &:= \hat{W}(X,P,0) = \hat{K}(X,P) - \frac{\gamma}{2}\,,
\end{align}
where the sampling radius \cite{runeson2019spin,runeson2020generalized} is
\begin{align}
    R^2 &= 2\sqrt{S+1}\label{eq:radius}\,.
\end{align}
We use the subscript ``scs" since $\hat{W}_{\rm scs}(X,P)$ has a close connection to the spin coherent state\cite{mannouch2020partially,mannouch2020partially2}. One can use the kernel and full sampling integration weight to express the ``closure relation" and arbitrary electronic operator
\begin{align}
    \hat{I} &= \int dXdP \rho_{\rm full}(X,P)\hat{W}_{\rm scs}(X,P)\label{eq:scsfullIdentiy}\,, \\
    \hat{B} &= \int dXdP \rho_{\rm full}(X,P)\hat{W}_{\rm scs}(X,P)\Tr{\hat{W}_{\rm scs}(X,P)\label{eq:scsfullMatrix}\hat{B}}\,.
\end{align}
The full sampling Spin-LSC method approximates the initial phase space distribution as the product of the full sampling integration weight times the SW transformation of $\rho_{\rm el}(0)$, which yields 
\begin{align}
    f_{\rm Spin-LSC,full}(\Gamma) &= \Tr{\hat{W}_{\rm scs}(X,P)\rho_{\rm el}(0)}\rho_{\rm full}(X,P)
\,,
\end{align}
and for $\rho_{\rm el}(0) = \ket{r}\bra{r}$
\begin{align}
    f_{\rm Spin-LSC,full}(\Gamma) &= \frac{1}{2}(X_r^2 + P_r^2 - \gamma)\rho_{\rm full}(X,P)\,.
\end{align}

There exists another initial sampling strategy that can also  express the trace of two electronic operators, ``closure relation" and arbitrary electronic operator, named focus sampling\cite{runeson2019spin,runeson2020generalized,mannouch2020partially,mannouch2020partially2},
\begin{align}\label{eq:WKernel}
    &\rho_{\rm foc}^{(m)}(X,P)= \frac{1}{\pi^S}\delta(X_m^2 + P_m^2 - \gamma - 2)\prod_{n\neq m}\delta(X_n^2 + P_n^2 - \gamma), \\
    & \Tr{\hat{B}\hat{C}} = \sum_m\int dXdP \rho_{\rm foc}^{(m)}(X,P) \nonumber \\
    &\times\Tr{\hat{W}_{\rm scs}(X,P)\hat{B}}\Tr{\hat{W}_{\rm scs}(X,P)\hat{C}}\,, \\
    &\hat{I} = \sum_m\int dXdP \rho_{\rm foc}^{(m)}(X,P)\hat{W}_{\rm scs}(X,P)\,,\label{eq:scsfocIdentity} \\
    &\hat{B} = \sum_m\int dXdP \rho_{\rm foc}^{(m)}(X,P)\hat{W}_{\rm scs}(X,P)\Tr{\hat{W}_{\rm scs}(X,P)\hat{B}} \label{eq:scsfocMatrix}.
\end{align}

The focus sampling Spin-LSC samples the initial electronic phase space variables on several circles of the same hypersphere surface as full sampling, rather than the entire hypersphere surface, which gives 
\begin{align}
    f_{\rm Spin-LSC,foc}(\Gamma) &= \sum_{m}\Tr{\hat{W}_{\rm scs}(X,P)\rho_{\rm el}(0)}\rho_{\rm foc}^{(m)}(X,P)
\,,
\end{align}
and for $\rho_{\rm el}(0) = \ket{r}\bra{r}$
\begin{equation}
    f_{\rm Spin-LSC,foc}(\Gamma) = \rho_{\rm foc}^{(r)}(X,P)\,.
\end{equation}

In general, Spin-LSC cannot sample the intra-electron correlation correctly except for the two-level systems. Here, we list an explicit example with wrong intra-electron correlation for any $S>2$ systems. Considering $\hat{O}_1 = \hat{O}_2 = \ket{2}\bra{3} + \ket{3}\bra{2}$, one has 
\begin{align}
    &\braket{\frac{\hat{O}_1\hat{O}_2 + \hat{O}_2\hat{O}_1}{2}} = 0 \,,\nonumber\\ 
    &\llangle \Tr{\hat{O}_1\hat{K}(X_0,P_0)}\Tr{\hat{O}_2\hat{K}(X_0,P_0)}\rrangle_{\rm Spin-LSC(full)}  \nonumber\\ &=\frac{R^4}{4(S+1)}(\frac{R^2}{S+2}-\gamma)=\frac{2\sqrt{S+1}}{S+2}-\frac{2}{S}(\sqrt{S+1}-1)\,,\nonumber\\ 
    &\llangle \Tr{\hat{O}_1\hat{K}(X_0,P_0)}\Tr{\hat{O}_2\hat{K}(X_0,P_0)}\rrangle_{\rm Spin-LSC(focus)} \nonumber\\
    &= \frac{\gamma^2}{2} = 2\frac{S+2-2\sqrt{S+1}}{S^2}\,. 
\end{align}
The proof of correct intra-electron correlation sampling for 2-level systems is listed in the Appendix A.

\subsection{Spin-PLDM and PLDM}

As we mentioned in the Sec.~II, partially linearized methods insert the ``closure relations" between the initial density matrix and forward/backward propagator\cite{mannouch2020partially,mannouch2020partially2,huo2011communication,hsieh2012nonadiabatic}. Spin-PLDM\cite{mannouch2020partially,mannouch2020partially2} inserts the ``closure relation" based on the SW transformation kernel Eq.~(\ref{eq:scsfullIdentiy},\ref{eq:scsfocIdentity}), which yields 
\begin{align}
    h_{\rm Spin-PLDM,full}(X_0,P_0) &= \rho_{\rm full}(X_0,P_0) \\
    h_{\rm Spin-PLDM,foc}(X_0,P_0) &= \sum_{m}\rho_{\rm foc}^{(m)}(X,P)\,.
\end{align}
The two versions of ``closure relation" give the full sampling Spin-PLDM and focus sampling Spin-PLDM, respectively. The sampling radius and ZPE parameter are identical to Eq.~(\ref{eq:radius},\ref{eq:ZPE}).

PLDM and PBME treat the initial electron density matrix and observable in the same fashion, i.e., mapping the initial electron density matrix and observable onto the excited state of fictitious harmonic oscillators and the pure creation--annihilation operator, respectively. In PLDM\cite{huo2011communication,hsieh2012nonadiabatic}, the inserted identity is the ``closure relation" of the coherent state of the harmonic oscillators. The ``closure relation" and related properties of the coherent states are 
\begin{align}
    &\Phi(X,P) = \frac{\exp{-\sum_{m=1}^S\frac{1}{2}(X_m^2+P_m^2)}}{(2\pi)^{S}}\,, \\
    &\hat{W}_{\rm cs}(X,P) = \hat{K}(X,P)\,, \\
    &\hat{I} = \int dXdP \Phi(X,P)\hat{W}_{\rm cs}(X,P)\label{eq:csIdentity}\,, \\
    &\hat{O} = \int dXdP \Phi(X,P)\hat{W}_{\rm cs}(X,P)\Tr{\hat{W}_{\rm cs}(X,P)\hat{O}}\label{eq:csMatrix}\,,
\end{align}
which corresponds to the zero ZPE parameter of Eq.~(\ref{eq:Wmatrix}). Therefore, the initial sampling of PLDM is given by
\begin{align}
    h_{\rm PLDM}(X_0,P_0) &= \Phi(X,P)\,.
\end{align}

The proof of Spin-PLDM full sampling, Spin-PLDM focus sampling, and PLDM satisfying the intra-electron correlation is straightforward. Using the property of ``closure relation" Eq.~(\ref{eq:scsfullIdentiy}/\ref{eq:scsfocIdentity}/\ref{eq:csIdentity}) and the operator expression via $\hat{W}_{\rm scs}(X,P)/\hat{W}_{\rm cs}(X,P)$ Eq.~(\ref{eq:scsfullMatrix}/\ref{eq:scsfocMatrix}/\ref{eq:csMatrix}) in the right-hand side of Eq.~(\ref{eq:IntraEParDef}), we can immediately obtain the left-hand side of Eq.~(\ref{eq:IntraEParDef}). Similar to the mLSC cases, the proof does not use the explicit form of $\rho_{\rm el}$, thus can also applied to arbitrary electronic operators.

\section{Discussion}

In this section, we further investigate the connection between intra-electron correlation and the short-time accuracy of population dynamics for a selection of chemical motivated models. In particular, this allows us to explain the order of short-time accuracy that has been observed in previous numerical benchmarks. The correct intra-electron correlations can improve the short-time accuracy of population dynamics from $\mathcal{O}(t^2)$ to $\mathcal{O}(t^3)$ for a general Hamiltonian with both real and imaginary off-diagonal matrix elements, from $\mathcal{O}(t^3)$ to $\mathcal{O}(t^4)$ for the atom-in-cavity models, and from $\mathcal{O}(t^5)$ to $\mathcal{O}(t^6)$ for spin-boson models. However, the correct intra-electron correlation cannot improve the short-time accuracy for scattering models, as methods with  correct and wrong intra-electron correlations are both accurate up to $\mathcal{O}(t^3)$.

We choose $\hat{O}_1$ and $\hat{O}_2$ as two generalized Gell-Mann matrices. The generalized Gell-Mann matrices $\hat{\Lambda}_{\mu}$ can be divided into three classes\cite{zhu2019generalized,lang2021generalized}, the diagonal class $\hat{\Lambda}_{\rm D}$, the real off-diagonal class $\hat{\Lambda}_{\rm R}$, and the imaginary off-diagonal class $\hat{\Lambda}_{\rm I}$ 
\begin{widetext}

\begin{equation}\label{eq:GGM}
    \hat{\Lambda}_\mu = \left\{
    \begin{aligned}
&\frac{1}{\sqrt{2}}(\ket{m}\bra{n} + \ket{n}\bra{m})\in \hat{\Lambda}_{\rm R} \quad \rm for \quad 1\le \mu \le S(S-1)/2, \quad 1\le n < m \le S\,,\\
&\frac{1}{\sqrt{2}i}(\ket{n}\bra{m} - \ket{m}\bra{n})\in \hat{\Lambda}_{\rm I} \quad \rm for \quad S(S-1)/2 < \mu \le S(S-1), \quad 1\le n < m \le S\,,\\
&\frac{1}{\sqrt{m(m+1)}}\sum_{n=1}^{m}(\ket{n}\bra{n} - m\ket{m+1}\bra{m+1})\in \hat{\Lambda}_{\rm D} \quad \rm for \quad S(S-1) < \mu \le S^2 - 1, \quad 1\le m < S\,.
\end{aligned}
\right.
\end{equation}
\end{widetext}

With the help of the classification of generalized Gell-Mann matrices, the commutation relations have the following properties
\begin{equation}
    \begin{aligned}
        &[\hat{\Lambda}_{\rm D},\hat{\Lambda}_{\rm R}]\Rightarrow\hat{\Lambda}_{\rm I} \\
        &[\hat{\Lambda}_{\rm D},\hat{\Lambda}_{\rm I}]\Rightarrow\hat{\Lambda}_{\rm R} \\
        &[\hat{\Lambda}_{\rm R},\hat{\Lambda}_{\rm R}]\Rightarrow\hat{\Lambda}_{\rm I} \\
        &[\hat{\Lambda}_{\rm I},\hat{\Lambda}_{\rm I}]\Rightarrow\hat{\Lambda}_{\rm I} \\
        &[\hat{\Lambda}_{\rm R},\hat{\Lambda}_{\rm I}]\Rightarrow\hat{\Lambda}_{\rm R}+\hat{\Lambda}_{\rm D} \,,
    \end{aligned}
\end{equation}
where $[\hat{\Lambda}_{\rm R},\hat{\Lambda}_{\rm I}]\Rightarrow\hat{\Lambda}_{\rm R}+\hat{\Lambda}_{\rm D}$ means that the commutator between an operator spanned by $\hat{\Lambda}_{\rm R}$ and an operator spanned by $\hat{\Lambda}_{\rm I}$ is spanned by $\hat{\Lambda}_{\rm D}$ and $\hat{\Lambda}_{\rm R}$.

In this section, we always choose $\hat{O}\in\hat{\Lambda}_{\rm D}$ because it is connected with the population dynamics.  Because of symmetry, when $\hat{O}_1$ and $\hat{O}_2$ belong to different classes of generalized Gell-Mann matrices, Eq.~(\ref{eq:IntraEleDef}) always holds (l.h.s. = r.h.s. = 0) for the methods considered in this article. Similarly, Eq.~(\ref{eq:IntraEleDef}) also holds (l.h.s. = r.h.s. = 0) when $\hat{O}_1 \neq \hat{O}_2$, and $\hat{O}_1,\hat{O}_2$ belong to $\hat{\Lambda}_{\rm R}$ or $\hat{\Lambda}_{\rm I}$ simultaneously.

\subsection{General Hamiltonian}
In the usual chemical motivated models, $\hat{V}$ and thus $\frac{\partial \hat{V}(\hat{x}_0)}{\partial \hat{x}_0}$ are spanned by $\hat{\Lambda}_{\rm D}$ and $\hat{\Lambda}_{\rm R}$, and hence $[\hat{O},\frac{\partial \hat{V}(\hat{x}_0)}{\partial \hat{x}_0}]$ is spanned by $\hat{\Lambda}_{\rm I}$. Therefore, Eq,~(\ref{eq:bare}) equals to zero and all the considered methods are accurate up to $\mathcal{O}(t^3)$ in these scenarios. However, there do exist certain cases in which $\hat{V}$ is spanned by all three types of generalized Gell-Mann matrices, for instance, systems with light induced conical intersections\cite{schiro2021quantum,zhou2020protocol} and systems with magnetic forces\cite{wu2021semiclassical,miao2019extension,wu2022phase}. In these cases, the methods with correct intra-electron correlation have higher short-time accuracy than the wrong one.

\subsection{Scattering models}
In this subsection, we consider the most common $\hat{V}$ spanned by $\hat{\Lambda}_{\rm D}$ and $\hat{\Lambda}_{\rm R}$. In order to see the role of intra-electron correlation, the fourth order time derivative of $\braket{\hat{O}(t)}_{\rm m}-\braket{\hat{O}(t)}$, and the time derivative of $F(\Gamma_t,x_t)$ are required, 

\begin{widetext}
\begin{equation}
    \frac{d}{dt}F(\Gamma_t,x_t) = i\Tr{\hat{F}(\Gamma_t)[\frac{\partial\hat{V}(x_t)}{\partial x_t},\hat{V}(x_t)]} - \Tr{\frac{\partial^2 \hat{V}(x_t)}{\partial x_t^2}\frac{p_t}{m}\hat{F}(\Gamma_t)}\,,
\end{equation}

\begin{equation}\label{eq:FourthDifference}
    \begin{aligned}
    &\frac{d^4}{dt^4}\braket{\hat{O}(t)}_{\rm m}|_{t=0} - \frac{d^4}{dt^4}\braket{\hat{O}(t)}|_{t=0} =  -\braket{\frac{1}{4m^2}\big[[\hat{O}(0),\frac{\partial^2\hat{V}(0)}{\partial\hat{x}_0^2}],\frac{\partial^2\hat{V}(0)}{\partial\hat{x}_0^2}\big]} -3\braket{\big[[\hat{O}(0),\frac{\partial\hat{V}(0)}{\partial \hat{x}_0}],\hat{V}(0)\big]\frac{\partial \hat{V}(0)}{\partial \hat{x}_0}}\frac{1}{2m} \\
    &-\braket{\big[[\hat{O}(0),\hat{V}(0)],\frac{\partial\hat{V}(0)}{\partial \hat{x}_0}\big]\frac{\partial \hat{V}(0)}{\partial \hat{x}_0}}\frac{1}{2m}  -3\braket{\frac{\partial \hat{V}(0)}{\partial \hat{x}_0}\big[[\hat{O}(0),\frac{\partial\hat{V}(0)}{\partial \hat{x}_0}],\hat{V}(0)\big]}\frac{1}{2m} -\braket{\frac{\partial \hat{V}(0)}{\partial \hat{x}_0}\big[[\hat{O}(0),\hat{V}(0)],\frac{\partial\hat{V}(0)}{\partial \hat{x}_0}\big]}\frac{1}{2m}  \\
    &-\braket{\frac{1}{2m}[\hat{O}(0),\frac{\partial\hat{V}(0)}{\partial \hat{x}_0}][\frac{\partial\hat{V}(0)}{\partial \hat{x}_0},\hat{V}(0)]} -\braket{\frac{1}{2m}[\frac{\partial\hat{V}(0)}{\partial \hat{x}_0},\hat{V}(0)][\hat{O}(0),\frac{\partial\hat{V}(0)}{\partial \hat{x}_0}]} \\
    &+\frac{1}{m}\int dx_0dp_0W_{\rm nuc}(x_0,p_0) \llangle\Tr\Big\{\hat{A}(\Gamma_0,t)\{\Tr{\hat{F}(\Gamma_0)\frac{\partial\hat{V}(x_0)}{\partial x_0}}\{3\big[[\hat{O},\frac{\partial\hat{V}(x_0)}{\partial x_0}],\hat{V}(x_0) \big]+\big[[\hat{O},\hat{V}(x_0)], \frac{\partial\hat{V}(x_0)}{\partial x_0}\big]\} \\
    &+[\hat{O},\frac{\partial\hat{V}(x_0)}{\partial x_0}]\Tr{\hat{F}(\Gamma_0)[\frac{\partial\hat{V}(x_0)}{\partial x_0},\hat{V}(x_0)]}
    \}\Big\}\rrangle  \,,
    \end{aligned}
\end{equation}

\end{widetext}
where we already ignored the trivial canceling terms at $t=0$, such as the expectation of single electronic operator and intra-electron correlation terms for $\hat{O}_1$, $\hat{O}_2$ spanned by different generalized Gell-Mann matrices classes. The explicit time derivative expressions of $\braket{\hat{O}(t)}$ and $\braket{\hat{O}(t)}_{\rm m}$ are listed in the Appendix B. The notation $\hat{V}(0) = \hat{V}(t=0)$ is equivalent to $\hat{V}(\hat{x}_0)$. The additional term $-\frac{1}{4m^2}\big[[\hat{O}(0),\frac{\partial^2\hat{V}(0)}{\partial\hat{x}_0^2}],\frac{\partial^2\hat{V}(0)}{\partial\hat{x}_0^2}\big]$ in the first line of Eq.~(\ref{eq:FourthDifference}) tells us that the correct intra-electron correlation cannot improve the short-time accuracy. This phenomenon has already been observed in the short-time population dynamics of Tully's models in Ref.~[\onlinecite{gao2020benchmarking}].

\subsection{Cavity-modified molecular dynamics}

Unlike the scattering models, atom-in-cavity models\cite{saller2021benchmarking} have vanishing $\frac{\partial^2 \hat{V}(\hat{x})}{\partial \hat{x}^2}$, which means the methods with correct intra-electron correlation are accurate up to at least $\mathcal{O}(t^4)$, while the methods with wrong intra-electron correlation are only accurate up to $\mathcal{O}(t^3)$. This corollary is confirmed in Ref.~[\onlinecite{saller2021benchmarking}]. In Ref.~[\onlinecite{saller2021benchmarking}], Saller \textit{et al.} reported that mLSC methods have tremendous improvement over Ehrenfest, PBME, and LSC-IVR. Specifically, mLSC$/\phi^1\phi^1$ (correct intra-electron correlation) outperforms mLSC$/\phi^2\phi^2$ and mLSC$/\phi^1\phi^2$ (wrong intra-electron correlation) in the three-level systems.

\subsection{Spin-boson models}

For simplicity, we consider one $1d$ boson bath in this subsection. The generalization to multi-dimensional multi-bosons bath is straightforward. For the spin-boson models \cite{leggett1987dynamics,caldeira1983quantum}, $U(\hat{x})$ is a quadratic polynomial of $\hat{x}$.  $\hat{V}(\hat{x})$ depends on $\hat{x}$ linearly, and $\frac{\partial\hat{V}}{\partial \hat{x}}$ is a diagonal position-independent operator. The above conditions give the following relations
\begin{equation}\label{eq:SpinBosonProperty}
\begin{aligned}
    [\hat{O}(t),\frac{\partial \hat{V}(t)}{\partial \hat{x}_t}]&=0 \,, \\
    \frac{\partial^m \hat{V}(t)}{\partial \hat{x}_t^m}&=0\,, \quad m > 1 \,, \\
    \frac{\partial^m U(t)}{\partial \hat{x}_t^m}&=0\,, \quad m > 2\,.
    \end{aligned}
\end{equation}

These specific properties make the higher order derivative of $\braket{\hat{O}(t)}$ and $\braket{\hat{O}(t)}_{\rm m}$ simplified drastically. The fourth order and fifth order time derivatives of $\braket{\hat{O}(t)}$ and $\braket{\hat{O}(t)}_{\rm m}$ at $t = 0$ coincide, and the difference of the sixth time derivative vanishes when the method can capture the correct intra-electron correlations. For the explicit time derivatives, see the Appendix B. Therefore, the methods with correct intra-elecron correlations are accurate up to at least $\mathcal{O}(t^6)$ for the population dynamics of the spin-boson models, while the methods with wrong intra-electron correlations are only accurate up to $\mathcal{O}(t^5)$. This coincides with previous reported numerical observations that all advanced methods can improve the accuracy in spin-boson models\cite{saller2019identity,runeson2019spin,mannouch2020partially,huo2011communication,hsieh2012nonadiabatic,hsieh2013analysis} (Notice that $S = 2$ in spin-boson models). Specifically, it gives an explanation why single Wigner mLSC outperforms PBME. The two methods are extremely similar, and the only difference is that the projector of the SEO subspace $\hat{\Pi}_1$ locates at the initial operator in PBME but at the observables in single Wigner mLSC.

The conclusion that correct intra-electron correlation can improve accuracy discussed above is not limited to spin-boson models. The time scale analysis also works on the other models which has the property Eq.~(\ref{eq:SpinBosonProperty}), for instance, FMO model \cite{gao2020simulating,kramer2018efficient,saller2019identity,saller2019improved,mannouch2020partially,runeson2019spin,runeson2020generalized} and Frenkel biexciton model \cite{gao2020simulating,gao2020benchmarking}. Our theoretical analyses in this subsection give an explanation to the previous numerical observations\cite{saller2019identity,saller2019improved,mannouch2020partially,runeson2019spin,runeson2020generalized} for FMO models, i.e., Spin-PLDM, PLDM, and mLSC/$\phi^1\phi^1$ sampling, give better short-time results than Ehrenfest, PBME, LSC-IVR, Spin-LSC, and mLSC/$\phi^2\phi^2$.

\section{Conclusions}
\label{sec:conclusions}
In this article, we have generalized the concept of intra-electron correlation, which has first been introduced in the context of the GDTWA \cite{zhu2019generalized,lang2021generalized}, to various mapping approaches. We have established rigorous connections between short-time accuracy and intra-electron correlation for various models. The correct intra-electron correlation can improve the short-time accuracy for Hamiltonians with both real and imaginary matrix elements, atom-in-cavity models, and spin-boson models, while it cannot for scattering models. We analytically prove that the Ehrenfest method, LSC-IVR, and PBME fail to correctly sample the intra-electron correlation even for two-level systems. Spin-PLDM, PLDM, three types of sampling for initial identity operator in mLSC, and mLSC with $\phi^2$ sampling for initial traceless operators can sample the intra-electron correlation faithfully for arbitrary $S$-level systems. While mLSC with $\phi^2$ sampling for initial traceless operators and Spin-LSC successfully sample the intra-electron correlation for two-level systems, they cannot sample the intra-electron correlation for $S$-level systems with $S > 2$. 

Our theoretical analyses give explanations on the previous numerical observations\cite{saller2019identity,saller2019improved,mannouch2020partially,runeson2019spin,runeson2020generalized,saller2021benchmarking,gao2020benchmarking} and they may provide a guideline for the development of future mapping approaches with increased accuracy. 
They also suggest that the benchmark results in Ref.~[\onlinecite{gao2020benchmarking}] on the two-level systems, which showed that mLSC/$\phi^1\phi^1$, mLSC/$\phi^1\phi^2$, and mLSC/$\phi^2\phi^2$ have similar accuracy, might be difficult to generalize to higher level systems. According to our analysis, mLSC/$\phi^1\phi^1$ and mLSC/$\phi^2\phi^1$ should be more accurate than mLSC/$\phi^1\phi^2$ and mLSC/$\phi^2\phi^2$ for higher-level systems in the short-time dynamics. Finally, we stress that the intra-electron correlation is only a measure for the accuracy of short-time dynamics, and methods that violate intra-electron correlations can outperform methods with correct intra-electron correlations in simulations of the long-time behavior. For instance, Spin-LSC, which incorrectly samples intra-electron correlations, has the best long-time accuracy among the methods considered in this article on FMO models~\cite{mannouch2020partially,mannouch2020partially2}.

\section*{Acknowledgments}
We thank Oriol Vendrell for helpful discussions.
We acknowledge support by Provincia Autonoma di Trento, the ERC Starting Grant StrEnQTh (Project-ID 804305), and Q@TN --- Quantum Science and Technology in Trento. 
P.H. has received funding from the Italian Ministry of University and Research (MUR) through the FARE grant for the project DAVNE (Grant R20PEX7Y3A). This project was funded within the QuantERA II Programme that has received funding from the European Union’s Horizon 2020 research and innovation programme under Grant Agreement No 101017733, by the European Union under NextGenerationEU, PRIN 2022 Prot. n. 2022ATM8FY (CUP: E53D23002240006), by the European Union under NextGenerationEU via the ICSC – Centro Nazionale di Ricerca in HPC, Big Data and Quantum Computing. Views and opinions expressed are however those of the author(s) only and do not necessarily reflect those of the European Union, The European Research Executive Agency, or the European Commission. Neither the European Union nor the granting authority can be held responsible for them.”

\section*{AUTHOR DECLARATIONS}
\subsection*{Conflict of Interest}
The authors have no conflicts to disclose.
\subsection*{Author Contributions}
\textbf{Haifeng Lang}: Conceptualization (lead); Methodology (lead); Investigation (lead); Writing – original draft (lead); Writing –
review \& editing (equal). \textbf{Philipp Hauke}: Conceptualization (supporting); Methodology (supporting); Funding acquisition
(lead);  Writing –review \&  editing (equal).

\section*{Data AVAILABILITY}
The data that support the findings of this study are available within the article.

\appendix

\section{Proof of intra-electron correlation for 2-level system}
In this Appendix, we will give the prove that traceless MMST with $\phi^2$ sampling for initial traceless operators and Spin-LSC can sample the correct intra-electron correlation for 2-level systems.

For two level system, the traceless operators $\hat{O}_1$ and $\hat{O}_2$ can only take the form 
\begin{align}
    \hat{O}_1 = s_{1,x}\hat{\sigma}_x +  s_{1,y}\hat{\sigma}_y + s_{1,z}\hat{\sigma}_z \,,\nonumber \\
    \hat{O}_2 = s_{2,x}\hat{\sigma}_x +  s_{2,y}\hat{\sigma}_y + s_{2,z}\hat{\sigma}_z
\end{align}
where all the coefficients of Pauli matrices are complex numbers. Using the algebra of the Pauli matrices and performing the integral in Eq.~(\ref{eq:mLSCQdIntra}), one can immediately obtain 
\begin{align}
    &\Tr{\hat{Q}_r\frac{\hat{O}_1\hat{O}_2 + \hat{O}_2\hat{O}_1}{2}} \equiv 0\,,\nonumber \\
    &\frac{1}{(2\pi)^S}\int d\Gamma_0 \phi^2\Tr{\hat{K}(X_0,P_0)\hat{Q}_r}\Tr{\hat{K}(X_0,P_0)\hat{O}_1}\nonumber \\
    &\Tr{\hat{K}(X_0,P_0)\hat{O}_2}\equiv 0\,,
\end{align}
which means $\phi^2$ approach for the traceless operator can sample the intra-electron correlation faithfully for 2-level systems. Again, $\hat{Q}_r$ can be replaced by arbitrary traceless operators without changing the conclusion.

For the Spin-LSC method applied to a 2-level system, it is straightforward to verify that
\begin{align}
    &\Tr{\rho_{\rm el}(0)\frac{\hat{O}_1\hat{O}_2 + \hat{O}_2\hat{O}_1}{2}} \equiv s_{1,x}s_{2,x} + s_{1,y}s_{2,y} + s_{1,z}s_{2,z}\,,\nonumber \\
    &\int d\Gamma \Tr{\hat{W}_{\rm scs}(X,P)\rho_{\rm el}(0)}\rho_{\rm full}(X,P) \nonumber \\
    &\times\Tr{\hat{K}(X,P)\hat{O}_1}\Tr{\hat{K}(X,P)\hat{O}_2} \nonumber \\
    &\equiv s_{1,x}s_{2,x} + s_{1,y}s_{2,y} + s_{1,z}s_{2,z}\,, \nonumber \\
    &\int d\Gamma \sum_{m}\Tr{\hat{W}_{\rm scs}(X,P)\rho_{\rm el}(0)}\rho_{\rm foc}^{(m)}(X,P) \nonumber \\
    &\times\Tr{\hat{K}(X,P)\hat{O}_1}\Tr{\hat{K}(X,P)\hat{O}_2} \nonumber \\
    &\equiv  s_{1,x}s_{2,x} + s_{1,y}s_{2,y} + s_{1,z}s_{2,z}\,,
\end{align}
where $\Tr{\rho_{\rm el}(0)} = 1$ is the only requirement of $\rho_{\rm el}(0)$. Due to the linearity of $\rho_{\rm el}(0)$ in the integral and quantum trace in above relations, one can immediately have
\begin{align}
    &\Tr{\hat{B}\frac{\hat{O}_1\hat{O}_2 + \hat{O}_2\hat{O}_1}{2}}  \nonumber \\
    \equiv &\int d\Gamma \Tr{\hat{W}_{\rm scs}(X,P)\hat{B}}\rho_{\rm full}(X,P) \nonumber \\
    \times&\Tr{\hat{K}(X,P)\hat{O}_1}\Tr{\hat{K}(X,P)\hat{O}_2} \nonumber \\
    \equiv &\int d\Gamma \sum_{m}\Tr{\hat{W}_{\rm scs}(X,P)\hat{B}}\rho_{\rm foc}^{(m)}(X,P) \nonumber \\
    \times&\Tr{\hat{K}(X,P)\hat{O}_1}\Tr{\hat{K}(X,P)\hat{O}_2} \,,
\end{align}
where $\hat{B}$ is an arbitrary operator. Therefore, both full sampling and focus sampling can successfully capture the intra-electron correlation for 2-level systems.

\section{Explicit time derivatives}
In this Appendix, we will list the explicit time derivative expressions of $\braket{\hat{O}(t)}_{\rm m}$ and $\braket{\hat{O}(t)}$ the general models, scattering models and spin-boson models.

The first, second, and third time derivatives of $\braket{\hat{O}(t)}$ and 
$\braket{\hat{O}(t)}_{\rm m}$ for the most general observable and Hamiltonian are
\begin{widetext}

\begin{equation}\label{eq:EOMExpectation1}
    \frac{d}{dt}\braket{\hat{O}(t)}_{\rm m} = -i\int dx_0dp_0W_{\rm nuc}(x_0,p_0)\llangle\Tr{\hat{A}(\Gamma_0,t)[\hat{O},\hat{V}(x_t)]}\rrangle\,,
\end{equation}

\begin{equation}\label{eq:EOMQuantum1}
    \frac{d}{dt}\braket{\hat{O}(t)} = -i\langle[\hat{O}(t),\hat{V}(t)]\rangle\,,
\end{equation}

\begin{equation}\label{eq:EOMExpectation2}
    \frac{d^2}{dt^2}\braket{\hat{O}(t)}_{\rm m} = \int dx_0dp_0W_{\rm nuc}(x_0,p_0)\{-\llangle\Tr{\hat{A}(\Gamma_0,t)\big[[\hat{O},\hat{V}(x_t)],\hat{V}(x_t)\big]}\rrangle -i\llangle\Tr{\hat{A}(\Gamma_0,t)[\hat{O},\frac{\partial\hat{V}(x_t)}{\partial x_t}\frac{p_t}{m}]}\rrangle\}\,,
\end{equation}

\begin{equation}\label{eq:EOMQuantum2}
    \frac{d^2}{dt^2}\braket{\hat{O}(t)} = -\langle\big[[\hat{O}(t),\hat{V}(t)],\hat{V}(t)\big]\rangle -i\langle[\hat{O}(t),\frac{\partial\hat{V}(t)}{\partial \hat{x}_t}\frac{\hat{p}_t}{2m} + \frac{\hat{p}_t}{2m}\frac{\partial\hat{V}(t)}{\partial \hat{x}_t}]\rangle\,,
\end{equation}

\begin{align}\label{eq:EOMExpectation3}
    \frac{d^3}{dt^3}\braket{\hat{O}(t)}_{\rm m} &=\int dx_0dp_0W_{\rm nuc}(x_0,p_0)\{ i\llangle\Tr{\hat{A}(\Gamma_0,t)\Big[\big[[\hat{O},\hat{V}(x_t)],\hat{V}(x_t)\big],\hat{V}(x_t)\Big]}\rrangle \nonumber \\ &-2\llangle\Tr{\hat{A}(\Gamma_0,t)\big[[\hat{O},\frac{\partial\hat{V}(x_t)}{\partial x_t}\frac{p_t}{m}],\hat{V}(x_t)\big]}\rrangle -\llangle\Tr{\hat{A}(\Gamma_0,t)\big[[\hat{O},\hat{V}(x_t)],\frac{\partial\hat{V}(x_t)}{\partial x_t}\frac{p_t}{m}\big]}\rrangle \nonumber\\
    &-i\llangle\Tr{\hat{A}(\Gamma_0,t)[\hat{O},\frac{\partial^2\hat{V}(x_t)}{\partial x_t^2}\frac{p_t^2}{m^2} + \frac{\partial\hat{V}(x_t)}{\partial x_t}\frac{F(\Gamma_t,x_t)-\partial_{x_t}U(x_t)}{m}]}\rrangle\}\,.
\end{align}

\begin{align}\label{eq:EOMQuantum3}
    \frac{d^3}{dt^3}\braket{\hat{O}(t)} &= i\langle\Big[\big[[\hat{O}(t),\hat{V}(t)],\hat{V}(t)\big],\hat{V}(t)\Big]\rangle -2\langle\frac{\hat{p}_t}{2m}\big[[\hat{O}(t),\frac{\partial\hat{V}(t)}{\partial \hat{x}_t}],\hat{V}(t)\big]\rangle -2\langle\big[[\hat{O}(t),\frac{\partial\hat{V}(t)}{\partial \hat{x}_t}],\hat{V}(t)\big]\frac{\hat{p}_t}{2m}\rangle \nonumber \\ 
    &-\langle\frac{\hat{p}_t}{2m}\big[[\hat{O}(t),\hat{V}(t)],\frac{\partial\hat{V}(t)}{\partial \hat{x}_t}\big]\rangle -\langle\big[[\hat{O}(t),\hat{V}(t)],\frac{\partial\hat{V}(t)}{\partial \hat{x}_t}\big]\frac{\hat{p}_t}{2m}\rangle
    +i\braket{[\hat{O}(t),\frac{\partial\hat{V}(t)}{\partial \hat{x}_t}]\frac{\partial U(t)}{\partial \hat{x}_t}}\frac{1}{m} \nonumber \\
    &+i\braket{[\hat{O}(t),\frac{\partial\hat{V}(t)}{\partial \hat{x}_t}]\frac{\partial\hat{V}(t)}{\partial \hat{x}_t}}\frac{1}{2m} +i\braket{\frac{\partial\hat{V}(t)}{\partial \hat{x}_t}[\hat{O}(t),\frac{\partial\hat{V}(t)}{\partial \hat{x}_t}]}\frac{1}{2m} -i\braket{\frac{\hat{p}_t}{2m}[\hat{O}(t),\frac{\partial^2\hat{V}(t)}{\partial \hat{x}_t^2}]\frac{\hat{p}_t}{m} } \nonumber \\
    &-i\braket{[\hat{O}(t),\frac{\partial^2\hat{V}(t)}{\partial \hat{x}_t^2}]\frac{\hat{p}_t^2}{4m^2} } -i\braket{\frac{\hat{p}_t^2}{4m^2}[\hat{O}(t),\frac{\partial^2\hat{V}(t)}{\partial \hat{x}_t^2}] }\,,
\end{align}

\end{widetext}
where $U(t) = e^{iHt}Ue^{-iHt}$ and $\hat{V}(t) = e^{iHt}\hat{V}e^{-iHt}$.

For $\hat{O}$ spanned by $\hat{\Lambda}_{\rm D}$ and $\hat{H}$ spanned by $\hat{\Lambda}_{\rm D}$ and $\hat{\Lambda}_{\rm R}$, the fourth-order time derivatives of $\braket{\hat{O}(t)}$ and 
$\braket{\hat{O}(t)}_{\rm m}$ at $t=0$ are

\begin{widetext}

\begin{align}\label{eq:EOMExpectation4t=0}
    \frac{d^4}{dt^4}\braket{\hat{O}(t)}_{\rm m}|_{t=0} &=\int dx_0dp_0W_{\rm nuc}(x_0,p_0)\{ \llangle\Tr\Big\{\hat{A}(\Gamma_0,0)\{\Big[\Big[\big[[\hat{O},\hat{V}(x_0)],\hat{V}(x_0)\big],\hat{V}(x_0)\Big],\hat{V}(x_0)\Big] \nonumber \\ 
    &-3\big[[\hat{O},\frac{\partial\hat{V}(x_0)}{\partial x_0}],\frac{\partial\hat{V}(x_0)}{\partial x_0}\frac{p_0^2}{m^2}\big] -3\big[[\hat{O},\frac{\partial^2\hat{V}(x_0)}{\partial x_0^2}],\hat{V}(x_0)\frac{p_0^2}{m^2}\big]-\big[[\hat{O},\hat{V}(x_0)],\frac{\partial^2\hat{V}(x_0)}{\partial x_0^2}\frac{p_0^2}{m^2}\big] \nonumber\\
    &-3\big[[\hat{O},\frac{\partial\hat{V}(x_0)}{\partial x_0}],\hat{V}(x_0) \frac{F(\Gamma_0,x_0)-\partial_{x_0}U(x_0)}{m}\big]-\big[[\hat{O},\hat{V}(x_0)], \frac{\partial\hat{V}(x_0)}{\partial x_0}\frac{F(\Gamma_0,x_0)-\partial_{x_0}U(x_0)}{m}\big]\nonumber \\
    &+\frac{1}{m}[\hat{O},\frac{\partial\hat{V}(x_0)}{\partial x_0}]\Tr{\hat{F}(\Gamma_0)[\frac{\partial\hat{V}(x_0)}{\partial x_0},\hat{V}(x_0)]}
    \}\Big\}\rrangle\}\,,
\end{align}

\begin{align}\label{eq:EOMQuantum4t=0}
    \frac{d^4}{dt^4}\braket{\hat{O}(t)}|_{t=0} &= \langle\Big[\Big[\big[[\hat{O}(0),\hat{V}(0)],\hat{V}(0)\big],\hat{V}(0)\Big],\hat{V}(0)\Big]\rangle +\braket{\frac{1}{4m^2}\big[[\hat{O}(0),\frac{\partial^2\hat{V}(0)}{\partial\hat{x}_0^2}],\frac{\partial^2\hat{V}(0)}{\partial\hat{x}_0^2}\big]} \nonumber \\  &-\braket{\frac{\hat{p}_0^2}{4m^2}\{3\big[[\hat{O}(0),\frac{\partial\hat{V}(0)}{\partial \hat{x}_0}],\frac{\partial\hat{V}(0)}{\partial \hat{x}_0}\big] +3\big[[\hat{O}(0),\frac{\partial^2\hat{V}(0)}{\partial \hat{x}_0^2}],\hat{V}(0)\big]+\big[[\hat{O}(0),\hat{V}(0)],\frac{\partial^2\hat{V}(0)}{\partial \hat{x}_0^2}\big]\}} \nonumber\\
    &-\braket{\{3\big[[\hat{O}(0),\frac{\partial\hat{V}(0)}{\partial \hat{x}_0}],\frac{\partial\hat{V}(0)}{\partial \hat{x}_0}\big] +3\big[[\hat{O}(0),\frac{\partial^2\hat{V}(0)}{\partial \hat{x}_0^2}],\hat{V}(0)\big]+\big[[\hat{O}(0),\hat{V}(0)],\frac{\partial^2\hat{V}(0)}{\partial \hat{x}_0^2}\big]\}\frac{\hat{p}_0^2}{4m^2}} \nonumber\\
    &-\braket{\frac{\hat{p}_0}{2m^2}\{3\big[[\hat{O}(0),\frac{\partial\hat{V}(0)}{\partial \hat{x}_0}],\frac{\partial\hat{V}(0)}{\partial \hat{x}_0}\big] +3\big[[\hat{O}(0),\frac{\partial^2\hat{V}(0)}{\partial \hat{x}_0^2}],\hat{V}(0)\big]+\big[[\hat{O}(0),\hat{V}(0)],\frac{\partial^2\hat{V}(0)}{\partial \hat{x}_0^2}\big]\hat{p}_0\}} \nonumber\\
    &+3\braket{\big[[\hat{O}(0),\frac{\partial\hat{V}(0)}{\partial \hat{x}_0}],\hat{V}(0)\big]\{\frac{\partial U(0)}{\partial \hat{x}_0}+\frac{\partial \hat{V}(0)}{\partial \hat{x}_0}\}}\frac{1}{2m} +\braket{\big[[\hat{O}(0),\hat{V}(0)],\frac{\partial\hat{V}(0)}{\partial \hat{x}_0}\big]\{\frac{\partial U(0)}{\partial \hat{x}_0}+\frac{\partial \hat{V}(0)}{\partial \hat{x}_0}\}}\frac{1}{2m} \nonumber \\
    &+3\braket{\{\frac{\partial U(0)}{\partial \hat{x}_0}+\frac{\partial \hat{V}(0)}{\partial \hat{x}_0}\}\big[[\hat{O}(0),\frac{\partial\hat{V}(0)}{\partial \hat{x}_0}],\hat{V}(0)\big]}\frac{1}{2m} +\braket{\{\frac{\partial U(0)}{\partial \hat{x}_0}+\frac{\partial \hat{V}(0)}{\partial \hat{x}_0}\}\big[[\hat{O}(0),\hat{V}(0)],\frac{\partial\hat{V}(0)}{\partial \hat{x}_0}\big]}\frac{1}{2m} \nonumber \\
    &+\braket{\frac{1}{2m}[\hat{O}(0),\frac{\partial\hat{V}(0)}{\partial \hat{x}_0}][\frac{\partial\hat{V}(0)}{\partial \hat{x}_0},\hat{V}(0)]}+\braket{\frac{1}{2m}[\frac{\partial\hat{V}(0)}{\partial \hat{x}_0},\hat{V}(0)][\hat{O}(0),\frac{\partial\hat{V}(0)}{\partial \hat{x}_0}]}
    \,.
\end{align}
\end{widetext}

For $\hat{O}$ spanned by $\hat{\Lambda}_{\rm D}$ and spin--boson Hamiltonian $\hat{H}$, the first and second time derivative of the force in the mapping approaches, the fourth, fifth, and sixth order time derivatives of $\braket{\hat{O}(t)}$ and $\braket{\hat{O}(t)}_{\rm m}$ are

\begin{widetext}

\begin{equation}
\begin{aligned}
    \frac{d}{dt}F(\Gamma_t,x_t) &= i\Tr{\hat{F}(\Gamma_t)[\frac{\partial\hat{V}(x_t)}{\partial x_t},\hat{V}(x_t)]}\,, \\
    \frac{d^2}{dt^2}F(\Gamma_t,x_t) &= \Tr{\hat{F}(\Gamma_t)\big[[\frac{\partial\hat{V}(x_t)}{\partial x_t},\hat{V}(x_t)],\hat{V}(x_t)\big]}\,,
    \end{aligned}
\end{equation}

\begin{align}\label{eq:EOMExpectation4}
    \frac{d^4}{dt^4}\braket{\hat{O}(t)}_{\rm m} &=\int dx_0dp_0W_{\rm nuc}(x_0,p_0) \llangle\Tr\Big\{\hat{A}(\Gamma_0,t)\{\Big[\Big[\big[[\hat{O},\hat{V}(x_t)],\hat{V}(x_t)\big],\hat{V}(x_t)\Big],\hat{V}(x_t)\Big] \nonumber \\ 
    &+2i\Big[\big[[\hat{O},\hat{V}(x_t)],\frac{\partial\hat{V}(x_t)}{\partial x_t}\frac{p_t}{m}\big],\hat{V}(x_t)\Big] +i\Big[\big[[\hat{O},\hat{V}(x_t)],\hat{V}(x_t)\big],\frac{\partial\hat{V}(x_t)}{\partial x_t}\frac{p_t}{m}\Big] \nonumber\\
    &-\big[[\hat{O},\hat{V}(x_t)], \frac{\partial\hat{V}(x_t)}{\partial x_t}\frac{F(\Gamma_t,x_t)-\partial_{x_t}U(x_t)}{m}\big]\}\Big\}\rrangle\,,
\end{align}

\begin{align}\label{eq:EOMQuantum4}
    \frac{d^4}{dt^4}\braket{\hat{O}(t)} &= \langle\Big[\Big[\big[[\hat{O}(t),\hat{V}(t)],\hat{V}(t)\big],\hat{V}(t)\Big],\hat{V}(t)\Big]\rangle +\braket{\big[[\hat{O}(t),\hat{V}(t)],\frac{\partial\hat{V}(t)}{\partial \hat{x}_t}\big]\frac{\partial U(t)}{\partial \hat{x}_t}}\frac{1}{m} \nonumber \\  &+i\langle\frac{\hat{p}_t}{2m}\{2\Big[\big[[\hat{O}(t),\hat{V}(t)],\frac{\partial\hat{V}(t)}{\partial \hat{x}_t}\big],\hat{V}(t)\Big] + \Big[\big[[\hat{O}(t),\hat{V}(t)],\hat{V}(t)\big],\frac{\partial\hat{V}(t)}{\partial \hat{x}_t}\Big]\} \rangle \nonumber\\ &+i\langle\{2\Big[\big[[\hat{O}(t),\hat{V}(t)],\frac{\partial\hat{V}(t)}{\partial \hat{x}_t}\big],\hat{V}(t)\Big] + \Big[\big[[\hat{O}(t),\hat{V}(t)],\hat{V}(t)\big],\frac{\partial\hat{V}(t)}{\partial \hat{x}_t}\Big]\} \frac{\hat{p}_t}{2m}\rangle \nonumber \\
    &+\braket{\big[[\hat{O}(t),\hat{V}(t)],\frac{\partial\hat{V}(t)}{\partial \hat{x}_t}\big]\frac{\partial \hat{V}(t)}{\partial \hat{x}_t}}\frac{1}{2m} + \braket{\frac{\partial \hat{V}(t)}{\partial \hat{x}_t}\big[[\hat{O}(t),\hat{V}(t)],\frac{\partial\hat{V}(t)}{\partial \hat{x}_t}\big]}\frac{1}{2m} \,,
\end{align}

\begin{align}\label{eq:EOMExpectation5}
    \frac{d^5}{dt^5}\braket{\hat{O}(t)}_{\rm m} &=\int dx_0dp_0W_{\rm nuc}(x_0,p_0) \llangle\Tr\Big\{\hat{A}(\Gamma_0,t)\{-i\Big[\Big[\Big[\big[[\hat{O},\hat{V}(x_t)],\hat{V}(x_t)\big],\hat{V}(x_t)\Big],\hat{V}(x_t)\Big],\hat{V}(x_t)\Big] \nonumber \\    &+3\Big[\Big[\big[[\hat{O},\hat{V}(x_t)],\frac{\partial\hat{V}(x_t)}{\partial x_t}\frac{p_t}{m}\big],\hat{V}(x_t)\Big],\hat{V}(x_t)\Big]+2\Big[\Big[\big[[\hat{O},\hat{V}(x_t)],\hat{V}(x_t)\big],\frac{\partial\hat{V}(x_t)}{\partial x_t}\frac{p_t}{m}\Big],\hat{V}(x_t)\Big] \nonumber \\
    &+\Big[\Big[\big[[\hat{O},\hat{V}(x_t)],\hat{V}(x_t)\big],\hat{V}(x_t)\Big],\frac{\partial\hat{V}(x_t)}{\partial x_t}\frac{p_t}{m}\Big]+3i\Big[\big[[\hat{O},\hat{V}(x_t)],\frac{\partial\hat{V}(x_t)}{\partial x_t}\frac{p_t}{m}\big],\frac{\partial\hat{V}(x_t)}{\partial x_t}\frac{p_t}{m}\Big] \nonumber \\
    & +i\frac{F(\Gamma_t,x_t)-\partial_{x_t}U(x_t)}{m}\{3\Big[\big[[\hat{O},\hat{V}(x_t)], \frac{\partial\hat{V}(x_t)}{\partial x_t}\big],\hat{V}(x_t)\Big] + \Big[\big[[\hat{O},\hat{V}(x_t)],\hat{V}(x_t) \big],\frac{\partial\hat{V}(x_t)}{\partial x_t}\Big] \} \nonumber\\
    &-\big[[\hat{O},\hat{V}(x_t)], \frac{\partial\hat{V}(x_t)}{\partial x_t}\{-\frac{\partial_{x_t}^2U(x_t)p_t}{m^2}+\frac{d}{dt}\frac{F(\Gamma_t,x_t)}{m}\}\big]\}\Big\}\rrangle\,,
\end{align}

\begin{align}\label{eq:EOMQuantum5}
    \frac{d^5}{dt^5}\braket{\hat{O}(t)} &= -i\langle\Big[\Big[\Big[\big[[\hat{O}(t),\hat{V}(t)],\hat{V}(t)\big],\hat{V}(t)\Big],\hat{V}(t)\Big],\hat{V}(t)\Big]\rangle +\langle\frac{\hat{p}_t}{2m}\{3\Big[\Big[\big[[\hat{O}(t),\hat{V}(t)],\frac{\partial\hat{V}(t)}{\partial \hat{x}_t}\big],\hat{V}(t)\Big],\hat{V}(t)\Big] \nonumber \\
    &+ 2\Big[\Big[\big[[\hat{O}(t),\hat{V}(t)],\hat{V}(t)\big],\frac{\partial\hat{V}(t)}{\partial \hat{x}_t}\Big],\hat{V}(t)\Big] + \Big[\Big[\big[[\hat{O}(t),\hat{V}(t)],\hat{V}(t)\big],\hat{V}(t)\Big],\frac{\partial\hat{V}(t)}{\partial \hat{x}_t}\Big]\} \rangle \nonumber \\&+ \langle\{3\Big[\Big[\big[[\hat{O}(t),\hat{V}(t)],\frac{\partial\hat{V}(t)}{\partial \hat{x}_t}\big],\hat{V}(t)\Big],\hat{V}(t)\Big] 
    + 2\Big[\Big[\big[[\hat{O}(t),\hat{V}(t)],\hat{V}(t)\big],\frac{\partial\hat{V}(t)}{\partial \hat{x}_t}\Big],\hat{V}(t)\Big] \nonumber\\
    &+ \Big[\Big[\big[[\hat{O}(t),\hat{V}(t)],\hat{V}(t)\big],\hat{V}(t)\Big],\frac{\partial\hat{V}(t)}{\partial \hat{x}_t}\Big]\}\frac{\hat{p}_t}{2m} \rangle
    +3i\langle\frac{\hat{p}_t^2}{m^2}\Big[\big[[\hat{O}(t),\hat{V}(t)],\frac{\partial\hat{V}(t)}{\partial \hat{x}_t}\big],\frac{\partial\hat{V}(t)}{\partial \hat{x}_t}\Big]\rangle \nonumber \\
    & -i\langle\frac{\partial U(t)}{\partial \hat{x}_t}\{3\Big[\big[[\hat{O}(t),\hat{V}(t)], \frac{\partial\hat{V}(t)}{\partial \hat{x}_t}\big],\hat{V}(t)\Big] + \Big[\big[[\hat{O}(t),\hat{V}(t)],\hat{V}(t) \big],\frac{\partial\hat{V}(t)}{\partial \hat{x}_t}\Big] \}\frac{1}{m}\rangle \nonumber\\
    & -i\langle\frac{\partial \hat{V}(t)}{\partial \hat{x}_t}\{3\Big[\big[[\hat{O}(t),\hat{V}(t)], \frac{\partial\hat{V}(t)}{\partial x_t}\big],\hat{V}(t)\Big] + \Big[\big[[\hat{O}(t),\hat{V}(t)],\hat{V}(t) \big],\frac{\partial\hat{V}(t)}{\partial \hat{x}_t}\Big] \}\frac{1}{2m}\rangle \nonumber\\
    & -i\langle\{3\Big[\big[[\hat{O}(t),\hat{V}(t)], \frac{\partial\hat{V}(t)}{\partial \hat{x}_t}\big],\hat{V}(t)\Big] + \Big[\big[[\hat{O}(t),\hat{V}(t)],\hat{V}(t) \big],\frac{\partial\hat{V}(t)}{\partial \hat{x}_t}\Big] \}\frac{\partial \hat{V}(t)}{\partial \hat{x}_t}\frac{1}{2m}\rangle \nonumber\\
    &-i\braket{[\frac{\partial\hat{V}(t)}{\partial \hat{x}_t},\hat{V}(t)]\big[[\hat{O}(t),\hat{V}(t)], \frac{\partial\hat{V}(t)}{\partial \hat{x}_t}\big]\frac{1}{2m}} -i\braket{\big[[\hat{O}(t),\hat{V}(t)], \frac{\partial\hat{V}(t)}{\partial \hat{x}_t}\big][\frac{\partial\hat{V}(t)}{\partial \hat{x}_t},\hat{V}(t)]\frac{1}{2m}} \nonumber \\
    &+\braket{\big[[\hat{O}(t),\hat{V}(t)], \frac{\partial\hat{V}(t)}{\partial \hat{x}_t}\big]\frac{\partial^2U(t)}{\partial \hat{x}_t^2}\frac{\hat{p}_t}{m^2}}\,,
\end{align}

\begin{align}\label{eq:EOMExpectation6}
    \frac{d^6}{dt^6}\braket{\hat{O}(t)}_{\rm m}|_{t=0} &=\int dx_0dp_0W_{\rm nuc}(x_0,p_0) \llangle\Tr\Big\{\hat{A}(\Gamma_0,t)\{-\Big[\Big[\Big[\Big[\big[[\hat{O},\hat{V}(x_0)],\hat{V}(x_0)\big],\hat{V}(x_0)\Big],\hat{V}(x_0)\Big],\hat{V}(x_0)\Big],\hat{V}(x_0)\Big] \nonumber \\ 
    &+8\frac{p_0^2}{m^2}\Big[\Big[\big[[\hat{O},\hat{V}(x_0)],\frac{\partial\hat{V}(x_0)}{\partial x_0}\big],\frac{\partial\hat{V}(x_0)}{\partial x_0}\Big],\hat{V}(x_0)\Big] + \frac{1}{m}\{-\frac{\partial U(x_0)}{\partial x_0}-\Tr{\hat{F}(\Gamma_0)\frac{\partial \hat{V}(x_0)}{\partial x_0}}\}\nonumber \\
    &\times\{6\Big[\Big[\big[[\hat{O},\hat{V}(x_0)],\frac{\partial\hat{V}(x_0)}{\partial x_0}\big],\hat{V}(x_0)\Big],\hat{V}(x_0)\Big]
    +3\Big[\Big[\big[[\hat{O},\hat{V}(x_0)],\hat{V}(x_0)\big],\frac{\partial\hat{V}(x_0)}{\partial x_0}\Big],\hat{V}(x_0)\Big]\} \nonumber \\
    &-\frac{1}{m}\Tr{\hat{F}(\Gamma_0)[\frac{\partial \hat{V}(x_0)}{\partial x_0},\hat{V}(x_0)]}\{4\Big[\big[[\hat{O},\hat{V}(x_0)],\frac{\partial\hat{V}(x_0)}{\partial x_0}\big],\hat{V}(x_0)\Big]
    +\Big[\big[[\hat{O},\hat{V}(x_0)],\hat{V}(x_0)\big],\frac{\partial\hat{V}(x_0)}{\partial x_0}\Big]\} \nonumber \\
    &-\frac{1}{m}\Tr{\hat{F}(\Gamma_0)\big[[\frac{\partial \hat{V}(x_0)}{\partial x_0},\hat{V}(x_0)],\hat{V}(x_0)\big]}\big[[\hat{O},\hat{V}(x_0)],\frac{\partial\hat{V}(x_0)}{\partial x_0}\big]
    \Big\}\rrangle\,,
\end{align}

\begin{align}\label{eq:EOMQuantum6}
    \frac{d^6}{dt^6}\braket{\hat{O}(t)}|_{t=0} &= \langle-\Big[\Big[\Big[\Big[\big[[\hat{O}(0),\hat{V}(0)],\hat{V}(0)\big],\hat{V}(0)\Big],\hat{V}(0)\Big],\hat{V}(0)\Big],\hat{V}(0)\Big] \nonumber \\ 
    &+8\frac{\hat{p}_0^2}{m^2}\Big[\Big[\big[[\hat{O}(0),\hat{V}(0)],\frac{\partial\hat{V}(0)}{\partial \hat{x}_0}\big],\frac{\partial\hat{V}(0)}{\partial \hat{x}_0}\Big],\hat{V}(0)\Big] + \frac{1}{m}\{-\frac{\partial U(0)}{\partial \hat{x}_0}-\frac{\partial \hat{V}(0)}{\partial \hat{x}_0}\}\nonumber \\
    &\times\{6\Big[\Big[\big[[\hat{O},\hat{V}(0)],\frac{\partial\hat{V}(0)}{\partial \hat{x}_0}\big],\hat{V}(0)\Big],\hat{V}(0)\Big]
    +3\Big[\Big[\big[[\hat{O}(0),\hat{V}(0)],\hat{V}(0)\big],\frac{\partial\hat{V}(0)}{\partial \hat{x}_0}\Big],\hat{V}(0)\Big]\} \nonumber \\
    &-\frac{1}{2m}[\frac{\partial \hat{V}(0)}{\partial \hat{x}_0},\hat{V}(0)]\{4\Big[\big[[\hat{O}(0),\hat{V}(0)],\frac{\partial\hat{V}(0)}{\partial \hat{x}_0}\big],\hat{V}(0)\Big]
    +\Big[\big[[\hat{O}(0),\hat{V}(0)],\hat{V}(0)\big],\frac{\partial\hat{V}(0)}{\partial \hat{x}_0}\Big]\} \nonumber \\
    &-\frac{1}{2m}\{4\Big[\big[[\hat{O}(0),\hat{V}(0)],\frac{\partial\hat{V}(0)}{\partial \hat{x}_0}\big],\hat{V}(0)\Big]
    +\Big[\big[[\hat{O}(0),\hat{V}(0)],\hat{V}(0)\big],\frac{\partial\hat{V}(0)}{\partial \hat{x}_0}\Big]\}[\frac{\partial \hat{V}(0)}{\partial \hat{x}_0},\hat{V}(0)] \nonumber \\
    &-\frac{1}{2m}\{\big[[\frac{\partial \hat{V}(0)}{\partial \hat{x}_0},\hat{V}(0)],\hat{V}(0)\big]\big[[\hat{O}(0),\hat{V}(0)],\frac{\partial\hat{V}(0)}{\partial \hat{x}_0}\big]+\big[[\hat{O}(0),\hat{V}(0)],\frac{\partial\hat{V}(0)}{\partial \hat{x}_0}\big]\big[[\frac{\partial \hat{V}(0)}{\partial \hat{x}_0},\hat{V}(0)],\hat{V}(0)\big]\}\rangle\,,
\end{align}

\begin{equation}
    \begin{aligned}
    &\frac{d^6}{dt^6}\braket{\hat{O}(t)}_{\rm m}|_{t=0} - \frac{d^6}{dt^6}\braket{\hat{O}(t)}|_{t=0} = \\
    &+\langle\frac{1}{2m}\{\big[[\frac{\partial \hat{V}(0)}{\partial \hat{x}_0},\hat{V}(0)],\hat{V}(0)\big]\big[[\hat{O}(0),\hat{V}(0)],\frac{\partial\hat{V}(0)}{\partial \hat{x}_0}\big]+\big[[\hat{O}(0),\hat{V}(0)],\frac{\partial\hat{V}(0)}{\partial \hat{x}_0}\big]\big[[\frac{\partial \hat{V}(0)}{\partial \hat{x}_0},\hat{V}(0)],\hat{V}(0)\big]\} \\
    & +\frac{1}{2m}\frac{\partial \hat{V}(0)}{\partial \hat{x}_0} \{6\Big[\Big[\big[[\hat{O},\hat{V}(0)],\frac{\partial\hat{V}(0)}{\partial \hat{x}_0}\big],\hat{V}(0)\Big],\hat{V}(0)\Big]
    +3\Big[\Big[\big[[\hat{O}(0),\hat{V}(0)],\hat{V}(0)\big],\frac{\partial\hat{V}(0)}{\partial \hat{x}_0}\Big],\hat{V}(0)\Big]\}   \\
         &+\frac{1}{2m}[\frac{\partial \hat{V}(0)}{\partial \hat{x}_0},\hat{V}(0)]\{4\Big[\big[[\hat{O}(0),\hat{V}(0)],\frac{\partial\hat{V}(0)}{\partial \hat{x}_0}\big],\hat{V}(0)\Big]
    +\Big[\big[[\hat{O}(0),\hat{V}(0)],\hat{V}(0)\big],\frac{\partial\hat{V}(0)}{\partial \hat{x}_0}\Big]\}   \\
    &+\frac{1}{2m}\{4\Big[\big[[\hat{O}(0),\hat{V}(0)],\frac{\partial\hat{V}(0)}{\partial \hat{x}_0}\big],\hat{V}(0)\Big]
    +\Big[\big[[\hat{O}(0),\hat{V}(0)],\hat{V}(0)\big],\frac{\partial\hat{V}(0)}{\partial \hat{x}_0}\Big]\}[\frac{\partial \hat{V}(0)}{\partial \hat{x}_0},\hat{V}(0)] \rangle \\
    &-\frac{1}{m}\int dx_0dp_0W_{\rm nuc}(x_0,p_0) \llangle\Tr\Big\{\hat{A}(\Gamma_0,t)\{ \Tr{\hat{F}(\Gamma_0)\big[[\frac{\partial \hat{V}(x_0)}{\partial x_0},\hat{V}(x_0)],\hat{V}(x_0)\big]}\big[[\hat{O},\hat{V}(x_0)],\frac{\partial\hat{V}(x_0)}{\partial x_0}\big]\\
    & +\Tr{\hat{F}(\Gamma_0)\frac{\partial \hat{V}(x_0)}{\partial x_0}}\{6\Big[\Big[\big[[\hat{O},\hat{V}(x_0)],\frac{\partial\hat{V}(x_0)}{\partial x_0}\big],\hat{V}(x_0)\Big],\hat{V}(x_0)\Big]
    +3\Big[\Big[\big[[\hat{O},\hat{V}(x_0)],\hat{V}(x_0)\big],\frac{\partial\hat{V}(x_0)}{\partial x_0}\Big],\hat{V}(x_0)\Big]\}  \\
    &+\Tr{\hat{F}(\Gamma_0)[\frac{\partial \hat{V}(x_0)}{\partial x_0},\hat{V}(x_0)]}\{4\Big[\big[[\hat{O},\hat{V}(x_0)],\frac{\partial\hat{V}(x_0)}{\partial x_0}\big],\hat{V}(x_0)\Big]
    +\Big[\big[[\hat{O},\hat{V}(x_0)],\hat{V}(x_0)\big],\frac{\partial\hat{V}(x_0)}{\partial x_0}\Big]\} \Big\}\rrangle      \,,
    \end{aligned}
\end{equation}

\end{widetext}
where we already neglect the trivial canceling terms. Using the properties of Wigner transformation Eq.~(\ref{eq:WignerExplicit},\ref{eq:WignerIdentity}) as well as the properties of the spin--boson model Eq.~(\ref{eq:SpinBosonProperty}), we obtain that the fourth and fifth order derivatives of $\braket{\hat{O}(t)}$ and $\braket{\hat{O}(t)}_{\rm m}$ at $t=0$ coincide, while the sixth order derivatives coincide if the method can sample the intra-electron correlation correctly.

\bibliography{main}

\end{document}